\documentstyle[11pt,newpasp,twoside]{article}
\input epsf
\def\edcomment#1{\iffalse\marginpar{\raggedright\sl#1\/}\else\relax\fi}
\reversemarginpar

\begin{document}
\title{X-ray Spectroscopy of Accretion Disks and Stellar Winds in X-Ray Binaries}
\author{Duane A.\ Liedahl and Patrick S.\ Wojdowski}
\affil{Department of Physics and Advanced Technologies,
Lawrence Livermore National Laboratory, P.O. Box 808, Livermore, CA 94550}
\author{Mario A.\ Jimenez-Garate and Masao Sako}
\affil{Columbia Astrophysics Laboratory and Department of Physics,
Columbia University, 538 West 120th Street, New York, NY 10027}

\begin{abstract}

From hot, tenuous gas dominated by Compton processes, to warm, photoionized
emission-line regions, to cold, optically thick fluorescing matter, accreting
gas flows in X-ray binaries span a huge portion of the parameter space
accessible to astrophysical plasmas.
The coexistence of such diverse states of material within small volumes
($10^{33}$--$10^{36}$ cm$^3$) leaves X-ray spectroscopists with a 
challenging set
of problems, since all such matter produces various X-ray spectral signatures
when exposed to hard X rays.
Emission-line regions in X-ray binaries are characterized by high 
radiation energy densities,
relatively high particle densities, and velocities $\sim1000$ km s$^{-1}$.
In this article, we describe some recent efforts to generate detailed 
X-ray line
spectra from models of X-ray binaries, whose aims are to reproduce spectra
acquired with the {\it ASCA}, {\it Chandra}, and {\it XMM-Newton} 
observatories.
With emphasis on the global nature of X-ray line emission in
these systems, the article includes separate treatments of high-mass
and low-mass systems, as well as summaries of continuum spectroscopy
and plasma diagnostics.

\end{abstract}

\section{Basic Concepts}

\subsection{X-Ray Binaries}

An X-ray binary (XRB) consists of a compact object---a
neutron star or black hole---that is accreting material
from  another (companion) star. XRBs comprise the brightest class of extrasolar
X-ray sources, and have been subjected to nearly four decades of study.
In the Galaxy, about 200 XRBs have been catalogued (van Paradijs 1995) and
grouped into two major subdivisions:
low-mass X-ray binaries (LMXB; with companion mass $M_c \la 1 M_{\sun}$)
and high-mass X-ray binaries (HMXB; $M_c \ga 10 M_{\sun}$), each with
a number of subclassifications in use in the current literature
(White 1989).

In the broadest astronomical context, XRBs owe their significance
to their status as products of exotic stellar evolution.
In a similar vein, XRBs
provide some of the best opportunities to observe the extreme
endpoints of stellar evolution,
in which issues such as dense matter and strong gravity come to the fore.
Accretion physics plays a role in all of these subject areas---we
learn about neutron stars and black holes by watching matter fall
onto (or into) them.
With the advent of spaceborne X-ray detectors with high spectral resolving
power comes new, highly-detailed information to inform our understanding
of mass transfer in XRBs. With respect to {\it Chandra} and {\it 
XMM-Newton}, it is too early
to provide a thorough review of the advances made possible by these 
new missions.
We will, therefore, keep the article somewhat general, and concentrate on
X-ray spectroscopy and modeling, and leave ``what-we-have-learned'' 
reviews for the future.
A series of review articles covering
most of the major topics associated with XRBs is contained in Lewin, 
van Paradijs,
\& van den Heuvel (1995).
A number of earlier review articles concerning the specifics of XRBs 
are in the literature.
Among them are Nagase (1989), White (1989), White \& Mason (1985),
and White, Swank, \& Holt (1983).

While the source of accretion material in both HMXBs and LMXBs is a 
nearby companion star,
the physical processes by which the material is transferred to the
collapsed star differ. For LMXBs, a gaseous disk mediates the transfer,
the disk being fed by Roche lobe overflow; for HMXBs, the UV field of
the early-type companion drives a wind through which the accretor moves,
sweeping up a small fraction of the wind through gravitational capture.
There are exceptions, and it is worth noting that
some HMXBs appear to be powered by Roche lobe
overflow, so that this classification scheme (LMXB/disk: HMXB/wind) is
not strictly adhered to.
The best studied XRBs typically have high accretion luminosities 
($10^{36}$--$10^{38}$
$\rm{erg~s^{-1}}$), owing to the ready availability of matter to accrete
(large $\dot{M}$) and to the compactness of the region in which kinetic energy
is converted into high-energy radiation ($L_x \propto \dot{M}/R_x$).

In LMXB accretion disks, the nearly circular quasi-Keplerian annuli
have projected velocities of $\sim10^3~\sin i 
~(M/M_{\sun})^{1/2}R_{10}^{-1/2}$ km s$^{-1}$,
for inclination angle $i$, where $R_{10}=R/(10^{10}$ cm).
In HMXBs, to the extent that we are observing reprocessing
in stellar winds, comparable velocities are expected. Large scale velocity
fields should broaden the lines, and, in some instances, shift them, as well.
With a few exceptions, the ability to measure the effects of velocity fields,
made possible by the high resolving power of the ${\it XMM-Newton}$ Reflection
Grating Spectrometer (RGS) and, especially, the {\it Chandra} High 
Energy Transmission
Grating (HETG), represents a new dimension in X-ray spectroscopy of XRBs,
and promises to provide valuable constraints on source models.

Even with the highest angular resolution available with X-ray
telescopes, XRBs cannot be spatially resolved, as they subtend
roughly 100 $\mu$arcsec at a distance of 1 kpc. The X-ray spectrum
measured at Earth is thus a composite of spectra formed
over a large volume and, presumably, over a wide range of physical
conditions. If we assume that the continuum is born pure at its site of
formation, then we can think of the interaction of the continuum
radiation with the accreting gas as deforming the continuum.
The magnitude of the continuum deformation varies wildly from
source to source, and, for a given source,
usually varies on observable time scales as well.
The spectral imprinting by circumsource material provides, in principle, a set
of handles that observers/analysts can use to construct a
physically and geometrically consistent model of the source.

\subsection{Emission Processes}

Continuum deformation in XRBs is the result of four main types of process:
continuum absorption, discrete emission and recombination continuum emission,
resonant scattering, and Compton scattering.
If local material intervenes with our line of sight
to the continuum source, then absorption may alter the shape
of the continuum, sometimes producing observable photoelectric edges.
The X-ray irradiated gas in the vicinity of the continuum source
radiates lines and continua characteristic of the local conditions.
This gas can also intercept continuum radiation at energies corresponding
to atomic level separations. A photon can be absorbed, and 
subsequently re-emitted at
nearly the same energy, but with a new direction. This can either add to
or subtract from the photon flux propagating toward the observer,
depending on the location of the reprocessing gas with respect to the
observer's line of sight to the continuum source.
Finally, in plasmas in which the abundant elements are stripped of
their electrons, Compton scattering can regulate the rate of energy
exchange between the gas and the radiation field, and can, depending
on the Compton depth and the electron temperature, produce substantial
modification to the spectrum.
The resulting composite spectrum is further modified by its
passage through the interstellar medium. Since most XRBs
lie near the Galactic plane at distances $\ga$ 1 kpc, this is significant,
in effect ``cutting off'' most sources at wavelengths longward of 20--30 \AA.

We emphasize emission-line spectroscopy in this article, and,
for the most part, restrict the discussion to
the 1--35 \AA\ band, since there are no observable
atomic features shortward of 1 \AA, and most XRBs cannot be
observed in the X-ray band at wavelengths longward of 35 \AA. The key
emission processes are briefly described below.

To generalize each of the reaction types, we denote by {\it core}
an atomic configuration and by {\it (core)~nl} the same configuration
modified by the addition of an electron described by the quantum
numbers $n$ and $l$. The symbols $e$ and $h\nu$ are used for free
electrons and photons, respectively.

\vskip 10pt
\noindent
{\it Electron-Ion Impact Excitation}
\begin{center}
$(core)~nl+e\rightarrow (core)~n^{\prime}l^{\prime} + e^{\prime}$
\end{center}
While electron temperatures in X-ray photoionized plasmas are too low 
(see \S1.3) to
effectively populate high-lying energy
levels by electron impact excitation (EIE), this process is often
the dominant one in transferring population between closely-spaced
levels. Important instances of EIE are the collisional de-population
of the energy level responsible for the so-called forbidden line
in He-like ions ($1s^2$-$1s2s~^3S_1$), and the populating of 
low-lying excited states
in L-shell and M-shell ions.
\vskip 10pt
\noindent
{\it Radiative Recombination}
\begin{center}
$core+e \rightarrow (core)~nl + h\nu$
\end{center}
The outgoing photons are distributed into a continuum---the
radiative recombination continuum (RRC)---above the recombination
edge ($E=\chi$), with a width $\Delta E \approx kT_e$. In an overionized plasma
$\Delta E/E \approx kT_e/\chi \ll 1$, which means that the 
recombination continuum
feature is narrow, and appears ``line-like.''
To date, the only identified RRC features
have been associated with recombination to the ground level of
H-like and He-like ions, but more often recombination
leaves an ion in an excited state.
\vskip 10pt
\noindent
{\it Dielectronic Recombination}
In this two-step process, the first step, radiationless capture, excites an
electron belonging to the initial-state ion---no photon is emitted.
\begin{center}
$(core)+e\rightarrow (core)^{\prime}~nl$
\end{center}
In X-ray photoionized plasmas, the core is excited by way of an 
intrashell transition,
such as $2s\rightarrow2p$.  The energy of the ion following capture lies
above the first ionization limit. The ion may, therefore, autoionize.
For example,
\begin{center}
$(core)^{\prime}~nl \rightarrow core +e^{\prime}$,
\end{center}
which is, in effect, an elastic scatter when combined with the first 
step. Alternatively, the
ion can emit a photon, leaving the ion with an energy below
the first ionization limit ({\it radiative stabilization}),
which completes the dielectronic recombination (DR),
for example,
\begin{center}
$(core)^{\prime}~nl \rightarrow (core)^{\prime}~n^{\prime}l^{\prime} +h\nu_o$.
\end{center}
DR is not an important process for K-shell ions, but can be the dominant
recombination mechanism for L-shell ions.
\vskip 10pt
\noindent
{\it Radiative Cascade}
\begin{center}
$(core)~nl\rightarrow (core)~n^{\prime}l^{\prime} +h\nu_o$
\end{center}
Subsequent to recombination into an excited level, the ion will decay 
in a series
of spontaneous radiative transitions, until it reaches the ground level.
At each step the ion may have access to several decay channels. The relative
probabilities attached to these various channels partly determine the 
ionic spectrum.
At high densities, however, EIE can interrupt the
cascade, possibly resulting in line ratios that differ noticeably 
from the low-density limit.
For ions with L-shell electrons, cascades following both RR and DR
must be considered. Since the post-DR stabilized state can have an 
excited core,
cascades that proceed in the presence of a ``frozen'' excited core
are possible, the spectrum of which is distinct from the RR cascade spectrum.
Since collisional excitation of high-lying states is exceptionally rare in
X-ray photoionized plasmas, radiative cascades following recombination
are of paramount importance, which can be contrasted to the case
of collisionally-ionized plasmas, where, in the context of line spectroscopy,
recombination processes are generally treated as details.

\vskip 10pt
\noindent
{\it Fluorescence}
\begin{center}
$1s^2 ~(nl)^N +h\nu \rightarrow 1s~ (nl)^N +e$
\vskip 5pt
$1s~ (nl)^N \rightarrow 1s^2~ (nl)^{N-1} + h\nu_K$
\end{center}
A two-step process.
Here the initial state ``core'' is written as the composite of a closed
$n=1$ shell plus $N$ additional bound
electrons. The first step is a K-shell photoionization. In the second step
one of the $N$ electrons
above the K shell fills the K shell, emitting a characteristic K 
photon. If the falling
electron originates in $n=2$ (L shell), then the photon is designated 
K$\alpha$.
The probability that the second step above occurs, rather than autoionization
\begin{center}
$1s~ (nl)^N \rightarrow 1s^2~ (nl)^{N-2} +e^{\prime}$
\end{center}
is called the fluorescence yield, which scales as $Z^4$.
Owing in part to the rapid scaling with $Z$, fluorescence lines
from iron are the most commonly observed. In principle, calculating
the fluorescence spectrum is a computationally intensive problem.
A full calculation of the K$\alpha$ fluorescent spectrum of a
few-times-ionized ion would require a determination of the level
population distribution of numerous low-lying energy levels,
consistent with the local electron density and
radiation field. The fact that many levels can have non-trivial population
densities in XRBs not only affects the exact distribution of fluorescent
lines but drastically increases the number of radiative and
autoionization transitions that need to be calculated. In practice,
however, these complications are ignored. In particular, the
redistribution of K$\alpha$ line emission cannot be resolved with
current instruments. The effect on fluorescent yields is not known
in general (cf., Jacobs et al.\ 1989).

\subsection{The Ionization Parameter}

The physical conditions in optically thin X-ray photoionized gas are,
for a given ionizing spectrum, conveniently parameterized by the
{\it ionization parameter}. Although there are a number of quantities
dubbed ``ionization parameter'' in the literature, most of what follows
is discussed in terms of $\xi$ (Tarter, Tucker, \& Salpeter 1969),
the use of which is motivated by consideration of the steady-state
equations of ionization equilibrium. Let $\beta_i$, $C_i$, and
$\alpha_{i+1}$ denote the photoionization rate (s$^{-1}$) of charge
state $i$, the collisional ionization rate coefficient ($\rm{cm^3 ~s^{-1}}$)
of $i$, and the recombination rate coefficient ($\rm{cm^3 ~s^{-1}}$)
of charge state $i+1$, respectively. The term $\alpha_{i+1}$ accounts
for all two-body recombination processes. In steady state the equations
of ionization equilibrium can be written
\begin{equation}
\beta_i n_i+n_e C_i(T_e)~ n_i =n_e  \alpha_{i+1}(T_e)~ n_{i+1},
\end{equation}
where the temperature dependence of the rate coefficients is
indicated.
Given the photoionization cross-section $\sigma_i$ and ionization
threshold energy $\chi_i$ of charge state $i$,
the photoionization rate for a
point source of ionizing continuum can be written
\begin{equation}
\beta_i =\frac{L_x}{r^2}~ \int_{\chi_i}^{\infty}dE~\frac{S_E(E)}{4\pi 
E}~\sigma_i(E),
\end{equation}
where $S_E$ is the spectral shape function, normalized on a suitable
energy interval. Denoting the integral in Eq.\ (2) by $\Phi_i$,
Eq.\ (1) becomes
\begin{equation}
\frac{L_x}{n_e r^2}~\Phi_i~ n_i + C_i (T_e)~n_i = \alpha_{i+1}(T_e) ~n_{i+1}.
\end{equation}
Let $\xi=L_x/n_e r^2$, which is called the {\it ionization parameter}. Then
\begin{equation}
\label{csd}
\frac{n_{i+1}}{n_i}=\frac{C_i(T_e)+\xi \Phi_i}{\alpha_{i+1}(T_e)}
\end{equation}
Equation (4) shows that an X-ray photoionized plasma with temperature $T_e$
is overionized with respect to collisional ionization
equilibrium, owing to the presence of the $\xi \Phi_i$ term.
A useful way to think about overionization is that
for a given level of ionization the plasma is cool compared to
the equilibrium temperature corresponding to collisional
ionization equilibrium (the limit $\xi \rightarrow 0$).
The simultaneous solution of the set of ionization equations (Eq.\ 4)
and the energy equation ($heating=cooling$) gives the $T_e(\xi)$ relation.
It is found that $C_i$ can usually be neglected for the photoionized 
plasmas of interest
to X-ray spectroscopists. Such plasmas are thus purely X-ray photoionized.

\subsection{X-Ray Photoionization Codes}

Thorough treatments of photoionization codes and their applications 
can be found in
Davidson \& Netzer (1979), Ferland et al.\ (1998), and Kallman \& 
McCray (1982).
In short, photoionization codes are used to determine the effect of a radiation
field on a gas of specified chemical composition, and the self-consistent
effect that passage through the gas has on the radiation field, including
the addition of local sources of radiation.
This entails, among other things, the determination of the charge 
state distribution and temperature
as functions of $\xi$. Maxwellian
electron distributions are assumed, but LTE is not. LTE is rarely applicable,
owing to the insufficiently high electron densities and the 
highly-diluted non-Planckian
radiation fields found in XRBs. Calculations are
performed in the spirit of {\it detailed level accounting},
whereby charge state fractions and level populations are calculated
by explicit inclusion of all relevant rates into and out of each
energy level. It is found
that highly-ionized ions exist over the approximate range 10--$10^4$ in
$\xi$ (c.g.s.\ units). Below that range recombination lines fall at 
energies below
the X-ray band, but inner-shell fluorescence line emission is still possible.
Above that range, ions are fully stripped, but hydrogen-like recombination
emission is still possible. Recombination spectra are formed primarily
in the approximate temperature range $10^5$-$10^6$ K.

\section{Spectral Diagnostics in X-Ray Photoionized Plasmas}

One of the themes of this article is the global nature of
line formation in XRBs. In other words, even simple models
of the geometric and physical distribution of gas in XRBs show that
lines form over regions comparable in size to the binary systems
themselves, often encompassing a broad range of physical conditions.
That the volume integral of the line emissivity determines the
observed line ratios and line profiles in a non-trivial way should
be kept in mind. Attempting to infer the run of physical conditions
in a source by solving the inverse problem using a set of line ratios
is problematic when working outside the framework of a particular
model of the source. Nevertheless, it is always useful to take
advantage of spectral features or line ratios that are, for example,
sensitive to temperature or density, in order to provide a starting
point for more detailed studies. In this section, a few of the available
discrete diagnostics are discussed.
A more detailed treatment of X-ray spectroscopy
of photoionized plasmas, with emphasis on atomic kinetics,
is given in Liedahl (1999).
A more general presentation of X-ray spectroscopy, including
discussions of the relevant atomic rates, can be found in Mewe (1999).

\subsection{Signatures of Dominance by X-Ray Photoionization}

Although X-ray line emission in XRBs is likely to originate
in photoionized gas, one still seeks explicit confirmation through
examination of spectra. After all, it is conceivable
that XRBs contain regions of, for example, shock-heated gas.
The indicator of photoionization dominance is recombination
dominance in the emission line spectra of high charge states.
Having discerned that the emission mechanism
is consistent with pure recombination, one can then appeal to
photoionization codes in order to assign a rough value of $\xi$, from
the $f_i(\xi)$ relationship, and $T_e$,
from the $T_e(\xi)$ relationship, both derived for the appropriate ionizing
continuum.
Fortunately, there are a few easily recognized spectral signatures
of recombination dominance, described below.

\vskip 10pt
\noindent
(1) {\it Presence of Narrow RRC.} As discussed in \S1.3, an X-ray photoionized
plasma is overionized, i.e., the electron temperature is much less than
the ionization temperature. This condition leaves recombination as the
dominant level populating mechanism (see \S1.2). X-ray lines produced
by radiative cascades following recombination are accompanied by 
(often) narrow RRC.
In the optically thin limit the ratio is a weak function only of temperature.
For H-like ions, for example, the ratio $I(1s$-$2p$)/I(RRC) is 
approximately 1.3
(Liedahl \& Paerels 1996). Since RRC shapes are determined by $T_e$, RRC
provide a relatively model-independent temperature diagnostic, and can,
in principle, be used to check the theoretical $T_e(\xi)$ relationship.

\vskip 10pt
\noindent
(2) {\it Large {\rm G} ratio in He-like ions.} The He-like $G$ ratio
is defined as the intensity ratio $(x+y+z)/w$, where $x$ and $y$ are
the intercombination lines $1s^2~^1S_0$--$1s2p~^3P_{2,1}$, $z$ is
the forbidden line $1s^2~^1S_0$--$1s2s~^3S_1$, and $w$ is the resonance
line $1s^2~^1S_0$--$1s2p~^1P_1$ (Gabriel \& Jordan 1969).
In collisional ionization equilibrium
$w$ is the brightest He-like line, while in photoionization equilibrium
in the low-density limit, $z$ is the brightest (Pradhan 1985; Liedahl 1999).
Thus a ``by-eye'' examination of the ratio $z/w$ may be sufficient
to discern photoionization/recombination
dominance.  At higher densities, however (see
next section), the blend $x+y$ can be brighter than $z$, so, to 
generalize, the sum
$x+y+z$ is used in the definition of $G$. In short, a large $G$ ($\approx 3$-4)
implies that recombination dominates, while $G \approx 1$ implies collisional
ionization (Bautista \& Kallman 2000; Porquet \& Dubau 2000). With velocity
fields of a few $\times~1000$ km s$^{-1}$, line broadening can cause 
$x$ and $y$ to blend
with $w$, so that $G$ cannot be trivially specified, and other
combinations of the lines may need to be used (Paerels et al.\ 2000).

\vskip 10pt
\noindent
(3) {\it Relatively weak or apparent absence of iron L-shell emission 
in presence
of K-shell emission from lighter elements.} Decades of solar
X-ray observations have familiarized spectroscopists with the fact that
bright iron L-shell emission dominates the line spectrum of 
cosmic-abundance plasmas
with temperatures in the $10^6$-$10^7$ K range (Phillips et al.\ 1982).
It was shown by Kallman et al.\ (1996), however, that under conditions of
overionization, iron L-shell emission is overwhelmed by $1s-np$ and 
RRC emission from
K-shell ions of,
for example, oxygen, neon, and magnesium. There are several reasons for this.
Perhaps the simplest way to look at it is to think of iron L-shell ions
as having unusually large collisional rate coefficients, which help to drive
the impressive line fluxes observed in collisionally ionized plasmas, but
do not come into play at all in photoionized plasmas. These calculations have
been borne out by the {\it ASCA} observation of
the HMXB Vela X-1 during eclipse (Nagase et al.\ 1994),
which shows a spectrum dominated by H-like and He-like ions.
In fact, Sako et al.\ (1999)
achieved satisfactory fits to the data using an otherwise detailed 
spectral model for
which iron L-shell spectra were entirely absent.
Recombination spectra also affect iron L-shell ratios (Liedahl et al.\ 1990)
but it appears that use of these ratios may have to wait for instruments of
higher sensitivity.

\subsection{The {\it R} Ratio in He-like Ions}

We have already remarked on the use of He-like ions to
discriminate between collisional ionization and photoionization.
They are also well known density diagnostics (Gabriel \& Jordan 1969).
The useful ratio is denoted by $R$, and is defined as $R=z/(x+y)$,
where $x$, $y$, and $z$ were defined in the previous section.
The $z$ line is a slow magnetic dipole transition, so that as
$n_{\rm e}$ increases, collisional depopulation of the
$1s2s~^3S_1$ level begins to compete with radiative decay.
The density at which the rates of these two depopulation mechanisms are equal
is called the critical density $n_{\rm crit}$.
The dominant sinks for the outgoing population flux are
the three $1s2p~^3P$ levels, two of which can decay to ground,
producing the $x$ and $y$ lines. Thus the ratio $R$
decreases with increasing $n_{\rm e}$.

The density dependence of
$R$ is well approximated by
$R=R_o~[1+(n_{\rm e}/n_{\rm crit})]^{-1}$, where
$R_o$ is the zero-density limit of $R$. The variation
of $R$ is greatest for the two orders of magnitude
in $n_{\rm e}$ centered on $n_{\rm crit}$. Below this
range, one would observe $R\approx R_o$, and assign an upper
limit to $n_{\rm e}$. Above this range, $R \approx 0$, and a lower
limit can be assigned.

The ground state transitions of He-like ions of
C, N, O, Ne, Mg, Si, S, Ar, Ca, and Fe
fall into the X-ray band.
Both $R_o$ and $n_{\rm crit}$ vary with $Z$ (Pradhan 1982),
and obtaining their theoretical values requires detailed
solutions of the rate equations (Bautista \& Kallman 2000; Porquet \& 
Dubau 2000).
In particular,
$n_{\rm crit}$ increases with $Z$, ranging from $\sim10^9$ cm$^{-3}$ 
for ${\rm C~ V}$
to $\sim10^{17}$ cm$^{-3}$ for Fe XXV. To the extent that the various
He-like spectra are present in a given set of data,
density information is available over a broad region of
$n_{\rm e}$--$\xi$ parameter space.

The prevalence of He-like ions in highly-ionized plasmas,
the relative simplicity of calculating He-like spectra, and their
sensitivity to plasma conditions
has made them the pre-eminent diagnostics in X-ray spectroscopy.
However, a complication that was noted early on in the study
of He-like spectra is now being revisited, one which may compromise
the use of the $R$ diagnostic in some systems, notably XRBs.
The collisional transitions that are crucial to the density sensitivity
of He-like ions,  $1s2s~^3S_1 \rightarrow 1s2p~^3P_{0,1,2}$ are all
electric dipole transitions, and can be driven equally well by photons.
The wavelengths corresponding to the energy separation of the
$1s2s~^3S_1$ and $1s2p~^3P_1$ levels, to take an example, are 1634 
\AA, 1273 \AA,
and 868 \AA, for O, Ne, and Si, respectively. A bright UV source in
the vicinity of the X-ray emission-line regions can thus have the
effect of mimicking the signature of high electron density.

The possible influence of UV radiation on $R$ was pointed out by
Gabriel \& Jordan (1969) for solar applications, where it was found
that only C V should be affected by photospheric UV emission, and
by Blumenthal, Drake, \& Tucker (1972), who emphasized that, for example,
the presence of a moderately hot white dwarf could drive the $R$ ratio
to low values. This has obvious implications for HMXBs, in which O and B
stars reside, and LMXBs, in which an accretion disk emits copious UV radiation.
It was argued by Kahn et al.\ (2001) that the low values of $R$ observed
in the isolated O supergiant $\zeta$ Pup are consequences of UV driving, rather
than high densities, consistent with expectations that
stellar winds are too tenuous for the collisional depopulation of
the $1s2s~^3S_1$ level to matter. However, if our expectations are 
that densities
{\it should} be high enough for the collisional mechanism to work, and a UV
source is also present, a genuine ``degeneracy'' occurs. This may be the
case in LMXBs (Liedahl et al.\ 1992) and some classes of cataclysmic 
variable. What is needed
is another X-ray density diagnostic, one that is not susceptible to
UV driving, and one for which the critical density lies in an interesting
part of parameter space. One such candidate is the Ne-like Fe XVII
17.10/17.05 ratio ($n_{\rm crit} \sim 10^{13}$ cm$^{-3}$),
which appears to be immune to the influence of few-eV
Planckian radiation fields (Mauche, Liedahl, \& Fournier, in
preparation).

\section{Photoionizing Continua in XRBs}

The X-ray spectra of XRBs are almost completely dominated by continuum
radiation, most of which is thought to be produced through conversion of the
gravitational potential energy of the accreted material into radiation.  Apart
from the nearly ubiquitous iron K complex,
detections of discrete line emission
have, prior to {\it ASCA}, been difficult to obtain,
owing to the limited sensitivity of X-ray
spectrometers, and partly to the small intrinsic equivalent widths of the
emission lines.  In the case of HMXBs, where most of the soft X-ray line
emission originates from an extended stellar wind, the mass-loss rates of the
companion stars are typically $\dot{M} \sim 10^{-6} ~M_{\odot}
~\rm{yr}^{-1}$, which implies photoelectric optical depths
$\tau_{\rm{pe}} \ll 1$ for most of the relevant charge states.
In LMXBs, where the lines are produced near the surface of an accretion disk,
the solid angle subtended by the disk at the continuum source is $\Delta
\Omega/4\pi \sim 0.1$--0.2.  Thus in either case only a small fraction
of the source X-ray luminosity leaves the system in the form of 
discrete X-ray emission.
Still, recent observations of HMXBs and LMXBs with high-sensitivity, 
high-resolution
spectrometers show discrete features that can be used to constrain models of
the structure of the accretion flow around compact objects (Brandt \& 
Schulz 2000;
Cottam et al.\ 2001).

Despite the large variety of XRBs, all of
them have one common property---they emit strong continuum radiation over
several decades in energy.  The X-ray luminosities of persistent
XRBs are intrinsically variable, and range anywhere from (typically) 
$L_x \sim 10^{36}
~\rm{erg~s}^{-1}$ during the low state to as high as the Eddington luminosity
of a $\rm{few} \times 10^{38} ~\rm{erg~s}^{-1}$.  In black hole X-ray novae, the
range in X-ray luminosities is more dynamic, with quiescent luminosities
as low as $L_x \sim 10^{30} ~\rm{erg~s}^{-1}$ (Garcia et al.\ 2000).  A large
fraction of XRBs, however, are persistent X-ray sources.

The intrinsic continuum shape and the location of the source determine the
ionization and thermal structure of the surrounding photoionized medium, and,
as such, are important for interpreting the associated emission/absorption
line features imprinted on the continuum spectrum.  On the other hand,
these structures do not depend on the physical
mechanism responsible for the production of the primary continuum radiation.
Therefore, the line spectrum cannot be used to infer {\it how} the continuum is
generated.

 From an observational point-of-view, the spectral shape of the continuum
radiation from XRBs is well-understood, in the sense that they can be
reproduced by relatively simple mathematical prescriptions.  In 
neutron star HMXBs, for
example, the X-ray continuum can usually be described by a single power
law with photon indices in the range $\Gamma$ = 1.2--2.0, with an exponential
cut-off at energies above $E_{\rm{c}} \ga 10$ keV, and $e$-folding
energies of $E_{\rm{f}} \sim 10$--30 keV (Nagase 1989).  A possible
physical interpretation of this spectrum is unsaturated Comptonization
of soft seed photons as they travel through a hot, fully-ionized medium.  When
the observed spectrum covers a wide enough range in energy, measurements of
the power-law slope and the cut-off energy provide simultaneous estimates of
the electron temperature and optical depth of the Comptonizing medium, as
demonstrated by Eardley, Lightman, \& Shapiro (1975) and Sunyaev \& Titarchuk
(1980) using the hard X-ray continuum spectrum observed in the HMXB Cygnus
X-1.  Most HMXBs contain pulsars, which makes the radiation field anisotropic
as well as time-dependent.  The illuminated circumsource medium, 
therefore, responds to
either the instantaneous flux or the time-averaged flux, depending on the gas
density and the pulse-period.  Electron cyclotron absorption and/or 
emission features are also
seen in the hard X-ray spectra of many HMXBs, and can be used to estimate the
magnetic field strength of the neutron  star (Tr\"umper et al.\ 1978; 
Wheaton et al.\ 1979)

In LMXBs, single-component continuum models usually do not provide adequate
fits to the data (see White, Nagase, \& Parmar 1995 and references therein).
The spectrum, instead, typically consists of a cut-off power law similar to
those seen in HMXBs, but with an additional blackbody component with 
temperatures
of $kT_{\rm bb} \sim 1 ~\rm{keV}$, the latter presumably originating from the
boundary layer between the accretion disk and the neutron star surface.  In
many cases, the blackbody component is observed to be more variable than the
Comptonized component, implying that it is not the source of seed photons for
Comptonization (White, Peacock, \& Taylor 1985).  There is also evidence for
the blackbody component to be point-like, whereas the harder Comptonized
spectrum originates from a more extended region (Church et al.\ 1998),
providing further support for the picture described above.  Mitsuda et al.\
(1984) demonstrated that the use of multi-colored blackbody models (as a
replacement for the Comptonized component) provide constraints on the inner
radius of the emitting region.  However, White, Stella, \& Parmar (1988) argue
that multi-colored blackbody emission models do not provide
adequate descriptions of the spectra of most LMXBs, and are, 
moreover, physically
implausible, since electron scattering should modify the shapes of the intrinsic
spectra (Shakura \& Sunyaev 1973).

The fact that the continuum shape can be parametrized by simple
phenomenological models makes it difficult to infer the physical mechanisms
responsible for continuum radiation production.  Various physical models
can produce continuum spectra of similar shape, and the data usually 
cannot be used to
distinguish between different models, especially when the data cover a
relatively narrow band pass (e.g., Vacca et al.\ 1987).  For example, a $kT
\sim 10 ~\rm{keV}$ bremsstrahlung spectrum cannot be distinguished from
Comptonized emission in a mildly Compton thick medium.
If each of the components are absorbed through different (or time varying)
column densities, the continuum parameters derived under the assumption of a
single absorbing column may provide purely empirical results that have little
to do with reality.  Sometimes, the degeneracy can be broken by studying the
temporal behavior of the continuum during the various spectral states, as well
as quasi-periodic oscillations (QPOs) in different energy bands.

Many theoretical models have been developed in an attempt to unify the global
spectrum and the temporal behavior of XRBs.  For disk accretion onto
weakly-magnetized neutron stars Lamb (1989), Lamb \& Miller (1995), and
Psaltis, Lamb, \& Miller (1995) proposed a physically self-consistent model,
which describes the spectra and the QPOs during the various states.  In this
model, the neutron star magnetosphere produces soft X-ray seed photons through
electron cyclotron emission, which are Comptonized within the magnetosphere
itself, as well as in a hot corona that forms around the magnetosphere.  More
recently, the development of various branches of self-consistent accretion
flow solutions such as advection-dominated accretion flows (ADAF; Ichimaru
1977; Narayan \& Yi 1995) that successfully describe the spectral behavior of
black hole XRBs (Tanaka \& Lewin 1995; Tanaka \& Shibazaki
1996), advection-dominated inflow/outflow systems (ADIOS;
Blandford \& Begelman 1999), and convection-dominated accretion flows (CDAF;
Quataert \& Gruzinov 2000) have provided us with deeper insights about the
global energetics of accretion flows around compact objects.

\section{X-Ray Spectroscopy of Stellar Winds in HMXBs}

Using the first X-ray satellite {\it Uhuru},
Schreier et al.\ (1972) showed that periodic
changes in the X-ray pulse frequency of Cen X-3 could
be used to establish the binarity of the system.
This observation, along with the finding that the optical
candidate of Cyg X-1 was a spectroscopic binary (Bolton 1971;
Webster \& Murdin 1972),
was a key to confirming suggestions that many of the bright
variable X-ray sources known at the time
were indeed close binaries (Shklovsky 1967 proposed that
Sco X-1 is a neutron star binary).
The {\it Uhuru} observations of Cen X-3 showed that during
X-ray eclipses the X-ray flux level did not go to zero but that there
was a residual flux, indicating the presence of extended matter,
or, in any case, another source of X rays in the system.
Cen~X-3 was soon identified in
the optical with an O-type star (Krzemi\'{n}sky 1974).  Early type
stars had already been known to have strong winds with mass loss rates
$\dot{M} \sim10^{-6}~M_\odot$ yr$^{-1}$ and velocities $v \sim1000$ km s$^{-1}$
(Morton 1967). Thus it was natural to attribute the residual
eclipse flux in Cen~X-3 to scattering or other reprocessing by a stellar wind.
Similar explanations were given for the residual flux observed in
other eclipsing X-ray binary pulsars identified with early-type stars,
such as Vela X-1, 4U 1538-522, SMC X-1, and LMC X-4.

The hard X-ray continuum spectrum of the residual eclipse radiation
was generally observed to have the same form as the uneclipsed flux in
the hard X-ray band.  Therefore, electron scattering, which is 
independent of photon energy for
small Compton $y$ parameters and photon energies such that $h\nu \ll m_{\rm
e}c^2$, was implicated as the reprocessing mechanism (e.g., Becker et 
al.\ 1978).
The residual flux observed
in eclipse is usually a few percent of the flux outside of eclipse,
implying scattering optical depths $\tau_{\rm e}$ of a few percent. To get an estimate
of the corresponding electron density, we let the
companion star set the linear scale ($R_\star\sim10R_\odot$), and
estimate $n_{\rm e}$ from $\tau_{\rm e}=n_{\rm e}R_\star\sigma_T$, where
$\sigma_T$ is the Thomson cross section.  This gives a density 
$\sim10^{10}$ cm$^{-3}$ in
HMXB winds, which is consistent with
the densities in the winds of isolated massive stars.

The wind provides a
natural source of accretion fuel to power the X-ray source.  For the
wind density derived above, the theory of Bondi \&
Hoyle (1944) predicts an accretion rate sufficient to power a neutron
star luminosity of approximately $10^{36}$ $v_{1000}^{-3}$ erg s$^{-1}$,
where $v_{1000}$ is the wind velocity in multiples of 1000 km s$^{-1}$.
Therefore, while it is possible that, in
some HMXBs, the compact object is powered by accretion directly from 
a wind of the
type found in isolated massive stars, those that lie at the high end
of the luminosity range are not easily accommodated. Either
the wind must be much different than in an isolated massive star wind
(for example, the wind velocity is much lower),
or the compact object must be fueled in some other manner, such as
Roche lobe overflow.

For a luminosity of $10^{36}$--$10^{38}$ erg s$^{-1}$, the density 
and length scale
described above imply $\xi$ values in the range $10^2$--$10^4$.
For such large ionization parameters, astrophysically abundant
elements are typically ionized up to the K shell (e.g., Kallman \&
McCray 1982).  If we note additionally that the electron temperatures
implied for such gas are $10^5$--$10^7$ K, then we can estimate line
luminosities.  For Si XIV Ly$\alpha$, for example, the recombination 
line luminosity is
\begin{equation}
L_{\rm line}=n_{\rm e} n_{\rm H}A_{\rm Si}~VF\alpha_{\rm eff}~E_{\rm line},
\end{equation}
where $E_{\rm line}$ is the line photon energy, $V$ is the volume of 
the emission region,
$F$ is the ionization fraction of fully stripped silicon,
$A_{\rm Si}$ is the silicon abundance relative to hydrogen, and 
$\alpha_{\rm eff}$ is
the effective rate coefficient for
recombinations that result in emission of a Si XIV Ly$\alpha$ photon.  If we
use the density and length scale from above, take the solar abundance 
of silicon,
assume that silicon is fully stripped ($F=1$), and $\alpha_{\rm eff}
=4.2\times 10^{-12}$ cm$^3$ s$^{-1}$ (for a temperature of
100 eV; Osterbrock 1989) then we find a value of approximately
$L_{\rm line} \approx10^{32}$ erg s$^{-1}$.  The continuum spectra
of X-ray pulsars are
typically flat (photon index of $-1$) up to approximately 10 keV, above
which they are cut off. So, the specific monochromatic luminosity of
the continuum is approximately ($L_X$/10\,keV).  In eclipse, since the
apparent luminosity is reduced to a few percent of that outside of
eclipse, the monochromatic luminosity becomes
$10^{33}$ erg s$^{-1}$ keV$^{-1}$.
This gives equivalent
widths $\sim100$ eV, which are easily observable with {\it
ASCA}.  With the benefit of hindsight, we see that
lines from highly-ionized gas might have been expected.
It was somewhat surprising
to many of us, however, when Vela~X-1 was observed over an eclipse,
and several emission lines
from H-like and He-like ions were observed (Nagase et al.\ 1994).

Before the launch of {\it ASCA} in 1993, X-ray spectroscopy provided
information on the winds of HMXBs mostly through studies of scattering
and absorption of the continuum.  The properties of the scattered continuum
are almost entirely independent of the physical state and chemical composition
of the scattering gas. This is useful, in that
studies of the X-ray scattered continuum ``find'' all of the gas, but
problematic, in that the electron scattered continuum conveys no
information on the physical state of the gas.
Absorption tends to probe regions of low ionization,
and provides a valuable complement to studies of emission lines (e.g.,
Haberl, White, \& Kallman 1989) from highly ionized regions.
The iron K emission line complex provided some constraints on the state of
gas in the pre-{\it ASCA} era but with proportional counters, the 6.4,
6.7, and 6.9 keV lines of neutral or near-neutral,
helium-like, and hydrogen-like iron, respectively were often blended.
The absence of detailed information regarding the state of the
highly-ionized gas has compromised
our ability to describe the wind physics
because the mechanism by which the wind is accelerated is likely to depend
on the local temperature and ionization state.

\subsection{Ionization of Stellar Winds by an X-ray Source}

In massive, isolated stars, winds are driven as UV photons from the
star transfer their outward momentum to the wind via absorption line
transitions.  Castor, Abbott, \& Klein (1975) showed that line 
transitions provide the
opacity to power winds in isolated, massive stars, and that the winds
obey the velocity law
\begin{equation}
v(R)=v_\infty~\biggl(1-\frac{R_\star}{R}\biggr)^\beta,
\end{equation}
for a given distance $R$ from the center of the wind donor,
where $R_\star$ is the stellar radius, and where $\beta=0.5$, 
assuming a UV point source.
By accounting for the finite size of the companion star,
Pauldrach, Puls, \& Kudritzki (1986) showed that values of $\beta$
in the range 0.7--1.0 are better representations of the wind kinematics.

Given a velocity profile, the wind density can be calculated by
applying the equation of mass continuity, assuming spherical symmetry:
\begin{equation}
n(R)=\frac{\dot{M}}{4\pi\mu m_p v(R) R^2},
\end{equation}
where $\mu$ is the gas
mass per hydrogen atom. We can estimate the density from this equation
and compare it to our earlier estimate  based upon inferred Compton depths.
For $\mu=1.4$,
\begin{equation}
n=2 \times 10^{10}~\biggr(\frac{\dot{M}}{\rm{10^{-6} 
~M_{\odot}~yr^{-1}}}\biggl)~
\biggl(\frac{v}{10^8~ \rm{cm~s^{-1}}}\biggr)^{-1}~
\biggl(\frac{R}{10^{12}~ \rm{cm}}\biggr)^{-2}~\rm{cm^{-3}},
\end{equation}
which provides some validation to expectations that a more or less normal
stellar wind is responsible for reprocessing the central X-ray continuum.

If a point source of X rays is introduced into this system such that
the distance of a gas parcel from the source is denoted by $r$,
then from Eqs.\ (6) and (7),  the ionization parameter of a smooth wind is
\begin{equation}
\xi=\frac{4\pi\mu m_p Lv_\infty }{\dot{M}}~\biggl(\frac{R}{r}\biggr)^2 ~
\biggl(1-\frac{R_{\star}}{R}\biggr)^\beta.
\end{equation}
Contours of constant $\log \xi$ for a representative set of parameters
are plotted in Fig.\ 1. With a strictly spherically symmetric wind,
i.e., ignoring the effects
of non-inertial forces in the co-rotating frame,
stellar rotation, etc.,
the line of centers is an axis of cylindrical symmetry.
Since recombination timescales near the
binary are short compared to the orbital period, this structure
is virtually fixed in the co-rotating frame, modulo intrinsic time variations in
the X-ray continuum flux. This type of configuration was first
studied by Hatchett \& McCray (1977).
Surfaces of constant $\xi$ are surfaces of fixed charge state distribution,
as well as constant temperature.

At large $r$ (and, therefore, large $R$) both the wind density and the
ionizing flux fall as $r^{-2} \approx R^{-2}$, and $\xi$ approaches
a constant value $\xi_0=4\pi \mu m_{\rm p} L_{\rm x} v_{\infty}/\dot{M}$, or,
numerically
\begin{equation}
\xi_0 \approx 50 ~\biggl(\frac{L_x}{10^{36}~\rm erg~s^{-1}}\biggr)~
\biggl(\frac{v_\infty}{10^8~{\rm cm~s^{-1}}}\biggr)
\biggl(\frac{\dot{M}}{10^{-6}~M_\odot~{\rm yr}^{-1}}\biggr)^{-1}
~\rm{erg~cm~s^{-1}}.
\end{equation}
More significantly, $\xi_0$ is also close to the values of
$\xi$ at points on the system midplane
($r=R$; $\xi$ on the midplane is exactly equal to $\xi_0$ for the limiting case
$\beta=0$).
This latter aspect provides a way to estimate a characteristic value
of $\xi$ for a particular system, since
most of the X-ray emission will originate in the binary system between the
companion and the X-ray source.

What Fig.\ 1 does not show, of course, is the $n_{\rm e}^2$ weighting needed
to calculate line emissivities $j(n_{\rm e},\xi)=n_{\rm e}^2 P(\xi)$, where
$P$ is the line power (c.g.s.\ dimensions erg cm$^3$ s$^{-1}$).
If the $\xi$ structure for a particular system is such that
a given ion can exist in regions near the companion star, then,
for lines emitted by this ion,
these denser regions of the irradiated stellar wind
will have the highest line emissivities. On the other hand,
the luminosities of lines formed at somewhat higher
$\xi$, produced away from the stellar surface, benefit from
the larger volumes $\Delta V_{\rm line}$ over
which they can form ($L_{\rm line} \sim j_{\rm line} \Delta V_{\rm line}$).
Moving away from the companion,
the competition between the falloff in density and the increase
in the volume over which lines form determines the weightings
given to each ion in producing the overall recombination
line spectrum.
The coupling
of models for the ionization structure and the wind velocity profile provides
the means by which to calculate this distribution in weightings,
called the {\it differential emission
measure distribution}, discussed in the next section.
It turns out, however, that $\xi_0$ is
a good indicator of the spectrum, in the sense that prominent
lines will be produced in regions characterized by values of
$\xi$ centered approximately on $\xi_0$.

\begin{figure}
\plotfiddle{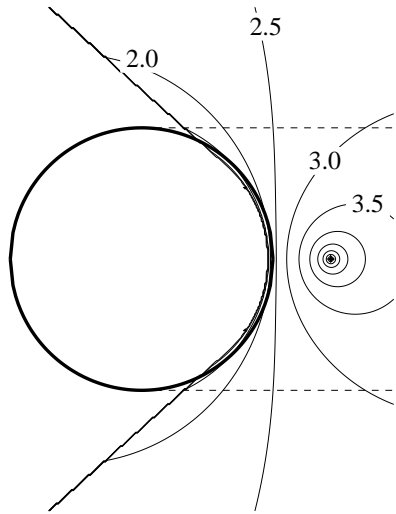}{2.9in}{-90.}{45.}{45.}{-200}{250}
\caption{Contours of $\log \xi$ in an HMXB with a smooth $\beta$-law 
wind (Eq.\ 9),
assuming the following parameters:
$L=10^{37}$ erg s$^{-1}$;
$\dot{M}=10^{-6}$ $M_{\sun}$ yr$^{-1}$;
$v_{\infty}=10^8$ cm s$^{-1}$; $\beta=0.8$; $R_{\star}=10~R_{\sun}$;
and $a=15~R_{\sun}$.
The line of centers is an axis of cylindrical symmetry.
Dashed lines delineate the part of the stellar wind occulted by
the companion star at orbital phase 0.0 (center of X-ray eclipse).
Diagonal lines tangent to the companion delineate the shadow cone---the
part of the wind shielded from direct radiation from the point 
X-ray source.}
\end{figure}
\noindent

The high level of ionization in the wind induced by the X-ray source
causes a serious complication in the simple picture presented above.
Studies of the effects of X rays on the wind-driving
mechanism have shown that the radiative force decreases as wind
material becomes increasingly ionized by the X-ray source (MacGregor \& Vitello
1982; Stevens \& Kallman 1990).
As predicted by Hatchett \& McCray (1977), this effect can be observed directly
through the orbital phase variation of resonance lines of UV P Cygni
profiles from ions such as Si~IV, C~IV, and N~V (Dupree et
al.\ 1980; Hammerschlag-Hensberge, Kallman, \& Howarth 1984; Vrtilek 
et al.\ 1997).
Stevens (1991), in one-dimensional numerical simulations,
calculated the effect of X-ray photoionization on
the wind dynamics in HMXBs, and was unable to find
outflowing wind solutions for X-ray luminosities larger than a few
times $10^{36}$ erg s$^{-1}$.

Despite the fact that winds in HMXBs may be very different from those
in isolated, massive stars, the density
distribution based upon the $\beta$-law (Eq.\ 6)
has been used as a convenient model.  For
example, Lewis et al.\ (1992) were able to model {\it Ginga} spectra
of Vela~X-1 using a Monte Carlo simulation of fluorescence and
scattering in a wind model for which
$\beta$ was set to 0.5 at distances greater than 0.3$R_\star$.
Closer to the stellar surface, a modification of the wind
structure, an exponential form with scale height $\sim0.1R_\star$
as proposed by Clark, Minato, \& Mi (1988) for Cen~X-3, was necessary 
to explain
the gradual eclipse transitions.
The physical processes of X-ray scattering and absorption are often
sufficient to describe the spectra seen with low-resolution
detectors.  However, an {\it ASCA} observation of Vela~X-1 over an eclipse
(Nagase et al.\ 1994) was dominated by recombination emission lines.  We describe
in the next section a method for calculating emission line spectra
for model winds.

\subsection{The DEM Approach}

When recombination is the dominant line production mechanism, the line
power can be written as
\begin{equation}
P_{\rm line}(\xi)= \frac{n_{\rm H}}{n_{\rm e}}~
AF_{i+1}(\xi)\alpha_{\rm eff}(\xi)~E_{\rm line},
\end{equation}
where $F_{i+1}$ is the ionic fraction of the recombining ion.
Recombination lines can be associated with particular values of $\xi$---each
line can be assigned a $\xi$ {\it of formation} $\xi_{\rm f}$.
These occur at the value of $\xi$ for which $P_{\rm{line}}(\xi)$ is a maximum,
equivalent to maximizing $F_{\rm {i+1}}(\xi) \alpha_{\rm {eff}}(\xi)$.
The {\it ASCA} eclipse spectrum of Vela X-1, for example, included 
emission from Ne IX
for which $\log\xi_{\rm f}=1.8$,
as well as Fe XXVI, for which
$\log\xi_{\rm f}=3.6$. The presence of bright emission lines from 
species of such
disparate $\xi_{\rm f}$ suggests that the reprocessing gas in Vela X-1
is composed of a non-trivial distribution in $\xi$. This comes as no 
great surprise,
based upon the discussion in the previous section, to the extent that the
simple model described there is a reasonable representation of the true matter
distribution.

To a first approximation, we can write
$L_{\rm line}=P_{\rm line}(\xi_{\rm f}) EM_{i+1}(\xi_{\rm f})$
where $EM_{i+1}=n_{\rm e}^2 V_{i+1}$ is called the {\it volume 
emission measure}.
Without corollary information that would allow an independent determination of
either the density or the emission volume, we can describe the ``amount''
of gas only in terms of the emission measure.
It is thus natural to plot the set $EM_{i+1}(\xi_{\rm f})$ vs.\ 
$\xi_{\rm f}$,
with each charge state represented by a point (Sako et al.\ 1999).
Dividing each $EM_{i+1}(\xi_{\rm f})$ by the width $\Delta \log \xi$,
an estimate of the width of $P(\xi)$, gives
an approximation to
the differential emission measure (DEM) distribution, the emission
measure per unit $\log \xi$, which is defined more formally below.
The significance of this is that the empirical DEM distribution
is relatively easy to construct, and can be compared to theoretical
predictions derived from geometrical models of the source. This has an
advantage over comparisons of spectra, since large-scale models of
XRBs often use highly schematic atomic models.
It is well-known, however, that recombination emission, especially
from H-like
ions, is efficient over a wide range in $\xi$ (Hirano et al.\ 1987; Ebisawa
et al.\ 1996). Therefore, constructing an empirical DEM distribution
in this way leaves us
with, at best, a crude approximation of the true, continuous DEM distribution.
With a geometrical model of the system, we can improve upon this
situation.

Electron scattering and fluorescence emissivities are proportional
to $n_{\rm e}L/r^2$ and $n_{i-1}L/r^2$, respectively. Therefore, by
substituting $n\xi$
for $L/r^2$, we see that the emissivities for scattering and
fluorescence can, just as for recombination lines, be written as $n_{\rm
e}^2$ times some function of $\xi$, which we denote by $P$ and call
the radiative power, a generalization of the line power discussed above.

From the definition of the volume emission measure $EM=\int dV~n_{\rm e}^2$,
we define the function $EM(\xi,v_\|)$ by limiting the volume of
integration to regions for which the ionization parameter
$\xi^\prime$ is less than the value $\xi$ and for which the radial
velocity $v_\|^\prime$ is less than the value $v_\|$.
The monochromatic luminosity of a photoionized gas distribution is then:
\begin{equation}
L_\nu=\int d\log\xi \int  dv_\|~
\biggl[\frac{\partial^2 EM(\xi,v_\|)}{\partial\log\xi~\partial v_\|}\biggr]~
P_\nu(\xi).
\end{equation}

With low to moderate resolving power, the Doppler broadening
and shifts of lines due to motion of
the wind are usually not detectable, and the $v_\|$
dimension in Eq.\ (12) is collapsed.
That is, we define $EM(\xi)$ by limiting the volume of
integration to the region with $\xi^\prime<\xi$, so that
\begin{equation}
L_\nu=\int d\log\xi~\biggl[\frac{dEM(\xi)}{d\log\xi}\biggr]~P_\nu(\xi).
\end{equation}
The quantities in brackets in Eqs\ (12) and (13) are DEM distributions.
The one-dimensional version is the only one that has seen use so far.

When calculating a large number of spectra,
where each is based upon a particular DEM distribution,
this representation, which splits the spatial and spectral components 
of local spectra,
is computationally efficient.
First, for a given ionizing radiation spectrum and chemical abundance set,
a library of $P_\nu(\xi)$ on a $\xi$-grid is compiled.
Then, for each wind model,
the DEM distribution is computed by summing the emission measure in
spatial cells into bins of $\log\xi$.  For a stellar wind which is
spherically symmetric, $\xi$ and $n_{\rm e}$ are symmetric about the
line of centers, and this symmetry significantly reduces the load of 
computing the
DEM distribution.  Where the wind models differ only by the luminosity
of the X-ray source or by an overall scaling, additional computational
efficiency is afforded, as the DEM distribution is changed only by
shifts along the
$\log\xi$ axis or by changes in the overall normalization,
and it is not necessary to repeat the summation over spatial cells 
(Wojdowski et al.\ 2000).

\subsection{Applications}

The DEM distribution for a radiatively driven wind depends not
only on the parameters of the wind ($\dot{M}$, $v_\infty$, and
$\beta$) and the luminosity of the X-ray source $L_{\rm x}$, but also on
the companion star radius $R_\star$ and the orbital separation
$a$. (Note that the dependence on $a$ is ``hidden'' in Eq. (9)
through the relation between $r$ and $R$.) Furthermore, the part of 
the wind which is
occulted at a given orbital phase depends on the orbital inclination $i$.  In
HMXBs where the compact object is a pulsar that is eclipsed during
its orbit, $R_\star$, $a$, and $i$ can be determined precisely using
Doppler pulse delays, Doppler shifts of optical lines from the
companion star, and the X-ray eclipse duration.  Furthermore, if the
distance to the system is known, $L_{\rm x}$ may also be constrained from
the uneclipsed flux.

Taking advantage of the fact that the system parameters
of Vela X-1 are  well constrained, Sako et al.\ (1999) computed DEM 
distributions
for a set of wind models,
assuming that the $\beta$-law is valid.
Emission spectra for these DEM distributions were calculated as
described above, and compared to the eclipse {\it ASCA}
spectrum. They found values of $\beta$ (0.79) and
$\dot{M}$ ($2.7 \times 10^{-7}$ $\rm{M_{\sun}~yr^{-1}}$) that
provided an accurate
reproduction of the recombination line spectrum. It is interesting
to compare, as a consistency check, the estimate of $\dot{M}$ that 
one would obtain from
Eq.\ (10). The value of $\xi_0$, using a plot
of $EM(\xi_{\rm f})$ vs.\  $\xi_{\rm f}$, as discussed in \S4.2,
can be estimated to be $\sim500$-1000. From Eq.\ (10),
this gives $\dot{M}\sim2.6$-5.2 $\times 10^{-7}$ $\rm{M_{\sun}~yr^{-1}}$,
in remarkably good agreement with the result based upon more detailed 
considerations.

The wind model used by Sako et al.\ does not,
however, predict fluorescent emission from low charge states,
which are observed in Vela X-1 at all orbital phases.
In order to obtain a good fit to the data, a set of features were
added to the fitting program to accommodate lines from
near-neutral Mg, Si, S, Ar, Ca, Fe, and Ni. A related problem is that
the highly-ionized wind model does not account for the intrinsic
near-neutral column density needed to attenuate the continuum flux.
Finally, and perhaps most glaringly, the derived $\dot{M}$ is an order of
magnitude smaller than the value derived from earlier studies (Dupree 
et al.\ 1980;
Sato et al.\ 1986), and considerably smaller than expectations based
upon wind mass loss rates from isolated early-type stars.
The problem can be summarized as follows: (1) the explanation of 
recombination emission
from K-shell ions requires a certain range of $\xi$ and a certain range
of DEM magnitudes, which leads to small $\dot{M}$;
(2) a model using a larger value of $\dot{M}$, while explaining the 
column density variation,
X-ray fluorescence, and UV evidence of low charge states, predicts
that the recombination line zones would closely surround
the neutron star, with dimensions so small that they would be occulted
by the companion star during eclipse.

Sako et al.\ showed that by abandoning the assumption that the wind is
smooth, this apparent discrepancy could be reconciled
(cf., Nagase et al.\ 1983; Sato et al.\ 1986). A picture of
the wind arises for which $\sim90\%$ of the wind's mass is in the form
of cool, dense clumps, but $\sim95\%$ of the irradiated wind volume within the
binary system conforms to the simple $\beta$-law/Hatchett-McCray model
used to fit the recombination emission.
This picture is able to bring the X-ray recombination spectrum,
the fluorescent line spectrum, the intrinsic column density,
and the presence of the low charge states observed in the UV into accord,
and yields an estimate of the total mass loss rate
($\dot{M}_{\rm tot} \sim 2 \times 10^{-6}$ $\rm{M_{\sun}~yr^{-1}}$) 
that agrees well
with earlier numbers.
Finally, this clumpiness may be related to short-term variability
($10^3$-$10^4$ s) in the X-ray luminosity through
collisions of dense wind structures with the neutron star.

In hot star winds, clumping is expected on theoretical grounds, and
is evidenced by the appearance of so-called ``discrete absorption components''
(DACs) in UV line profiles.  On a timescale of a few
days, DACs may appear in the absorption trough of
a UV P Cygni profile, which subsequently begin to narrow and move to the
blue (e.g., Kaper et al.\ 1996).  X rays are observed from isolated O
and B stars (e.g., Seward et al.\ 1979), whose origin has been attributed
to shocks in radiatively driven winds (Lucy 1982).
Line-driven stellar winds are subject to several instabilities
(MacGregor, Hartmann, \& Raymond 1979; Lucy \& White 1980;
Owocki \& Rybicki 1984), such as the Rayleigh-Taylor
instability, which will produce clumps with sizes $\sim10^{11}$ cm
(Carlberg 1980). Thus it is not entirely surprising that
a clumpy wind is invoked to explain the X-ray spectrum of Vela X-1.
The relation between clumping in a normal wind and clumping in
an X-ray irradiated wind is not established, however.
While the characteristics of the clumps in the model are reasonable,
they are more or less {\it ad hoc}, do not explain their origin or
survivability, and do not accommodate a clump
population that is likely to possess a distribution of sizes and densities.

Liedahl et al.\ (2000) speculate that the clumped component of the wind, since
it is characterized by small values of $\xi$, has sufficient opacity
to the companion's UV field to permit normal radiative driving.
While the diffuse wind is too highly ionized to be driven by UV radiation,
the clumps, which contain most of the wind mass, may mediate
the transfer of momentum to the outflow, with the result that
the hotter, diffuse wind is ``pushed'' outward, thereby
imparting the $\beta$-law velocity distribution to the recombination emission
line regions. Alternatively,
the hotter, more tenuous wind component may originate from the evaporation
of material off the clump surfaces, which then leaves it tied to
the clump velocity distribution.

Wojdowski et al.\ (2001) applied a similar DEM analysis of the
{\it ASCA} spectrum of Cen X-3.
Unlike Vela X-1, for which a wind terminal velocity has been measured
(Dupree et al.\ 1980),  $v_\infty$ for Cen X-3 is not known.
From Eq.\ (7) it can be seen
that the local density, hence DEM, scales with $\dot{M}/v_\infty$. Thus
the ratio $\dot{M}/v_\infty$ must be used as a fitting parameter, rather
than $\dot{M}$.
Also by contrast to Vela X-1, it is already known that any clump population
should have a negligible effect on the Cen X-3 X-ray spectrum, as evidenced
by the near constancy with orbital phase of the
equivalent width of the fluorescent iron K$\alpha$ complex (Ebisawa 
et al.\ 1996). In other
words, were the wind highly clumped and extended, a dramatic increase 
in the Fe K$\alpha$
equivalent width would accompany eclipse of the compact X-ray source. 
Vela X-1, for
example, {\it does} show a large-amplitude variability in the Fe 
K$\alpha$ equivalent width
(Sato et al.\ 1986). Therefore, it is assumed from the outset that
the wind in Cen X-3 is smooth, and that, for a given $v_{\infty}$,
the best-fit $\dot{M}$ pertains to the total mass-loss rate of the companion.

The value of $\dot{M}/v_\infty$ derived from the
Wojdowski et al.\ analysis of the {\it ASCA} Cen X-3 spectrum is $1.6 \times
10^{-9}~\rm{M_{\sun}~yr^{-1}}~(\rm{km~s^{-1}})^{-1}$,
which is comparable to that obtained for isolated stars
of similar spectral class. It is also comparable
to the value found for Vela~X-1,
if the clumped component's contribution to  $\dot{M}$ is included.
While the wind mass loss rates in Cen X-3 and Vela X-1
may be similar, the characters of the winds are quite different.
This is demonstrated by Fig.\ (2), which shows a comparison of the
DEM distributions obtained from fits to the {\it ASCA} data.
The apparent difference in
the total emission measures (the integrals of the
DEM distributions) is not real, and results simply
from the omission of the DEM distribution for the (off-scale) 
lower-$\xi$ region.
Indeed, the total emission measure in a spherically symmetric wind is
$EM \sim (\dot{M}/v_{\infty})^2/m_{\rm p}^2 R_{\star}$, in which a
substantial difference between the two systems is not to be found.
A reliable calculation of the distribution for $\log \xi <1.5$ would require a
thorough analysis of the line complexes from fluorescing
material, which is impossible without much higher spectral resolving power.
Above $\log\xi =1.5$, however, the difference between the two curves is real.
The relatively small DEM magnitudes in Vela X-1 reflect
the fact that most of the wind mass is ``locked up''
in dense clumps.

It is well known that X-ray irradiated gas is prey to thermal instabilities
near the range of ionization parameter relevant to X-ray recombination
line emission (see \S5.5). In this context the dimensionless {\it 
pressure ionization
parameter} ($\Xi=L/4\pi r^2 nkTc$ for a point X-ray source) is more useful
than $\xi$, which is sometimes called the {\it density ionization parameter}.
The instability criterion is given by $d\ln T/d\ln \Xi <0$,
often satisfied for a range of $\Xi$ in X-ray photoionized
gases, depending on the spectral shape function. Inside this range
($-1 \la \log \Xi \la 1$; Krolik, McKee, \& Tarter 1981),
both a dense, cool phase and a hot, tenuous phase may coexist
(i.e., there are two values of $\xi$ for a given $\Xi$ in this range).
To estimate $\Xi$ for winds in HMXBs, we use
the continuity equation (Eq.\ 7) to substitute for $n$ in the definition
of $\Xi$, which gives
\begin{equation}
\Xi=\frac{\mu m_{\rm p} Lv_{\infty}}{kT \dot{M}c}~\biggl(\frac{R}{r} \biggr)^2
~\biggl(1-\frac{R_{\star}}{r} \biggr)^{\beta}.
\end{equation}
As we did earlier for $\xi$, we can define a characteristic value of
$\Xi$ according to
\begin{equation}
\Xi_0=\frac{\mu m_{\rm p} L}{kTc ~(\dot{M} /v_{\infty}) }
=0.9~\frac{L_{36}}{T_6~(\dot{M}/v_{\infty})_{-9}},
\end{equation}
where $L$ is given as a multiple of $10^{36}$ erg s$^{-1}$,
$T$ as a multiple of $10^6$ K, and $\dot{M} /v_{\infty}$ as a multiple of
$10^{-9}~\rm{M_{\sun}~yr^{-1}}~(\rm{km~s^{-1}})^{-1}$.
For Vela X-1, with $(\dot{M} /v_{\infty})_{-9}=1.2$ for the total 
mass loss rate,
and $T_6\approx1$, we find $\log \Xi_0=0.5$, which is consistent with 
the two-phase
gas, possibly explaining the clumping. For Cen X-3, with its much higher X-ray
luminosity,  we find $\log \Xi_0=1.8$, above the range of $\Xi$ for 
which the thermal instability
applies---no cool phase is allowed. This is a crude argument, but it is
consistent with the X-ray spectra, and suggests
that further work into the formation and survivability
of clumped winds in HMXBs may show that thermal instabilities play
an important role.

\begin{figure}
\plotfiddle{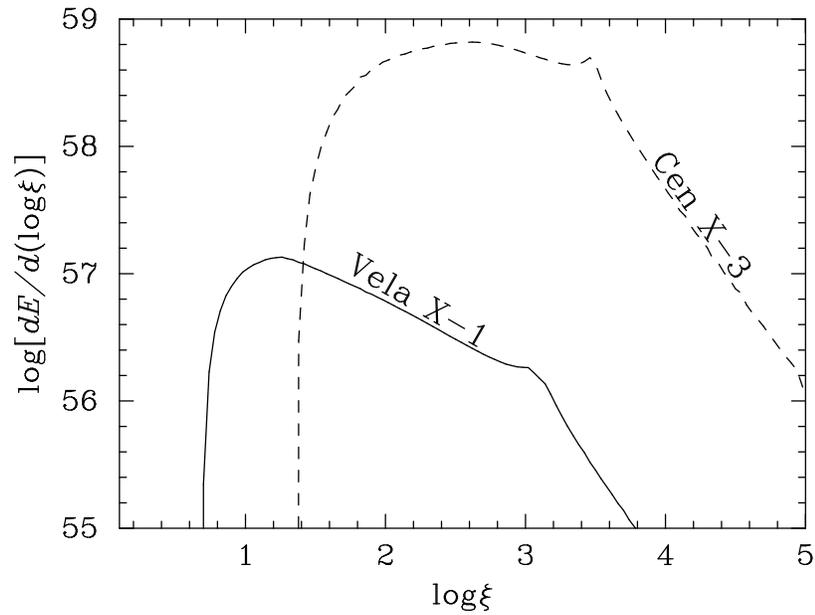}{2.9in}{-90.}{45.}{45.}{-170}{250}
\caption{Differential emission measure distributions of Vela~X-1
(Sako et al.\ 1999) and Cen~X-3 (Wojdowski et al.\ 2001).
For Vela X-1, the contribution from low-$\xi$ fluorescing material,
which would cause an upturn in the DEM curve moving to lower $\xi$,
is not included. The substantial difference in the DEM magnitudes
for $\log\xi >1.5$ illustrates the distinction between a smooth wind (Cen X-3)
and a clumped wind (Vela X-1). The vertical axis is scaled in units
of cm$^{-3}$.}
\end{figure}

\subsection{Physical Simulations}

The interplay of the various physical processes in HMXB winds is complex,
and computational simulations that include hydrodynamics, the effects of
radiation on the thermal balance,  and gravitational and non-inertial forces
are necessary to determine the resulting structure and kinematics.
As these simulations are necessarily
inexact, it is desirable to test them by predicting how these
simulated wind models would appear when observed with real
observatories and comparing these predictions with observational data.
In principle, simulations could be improved by comparing predicted
spectra to observed spectra, identifying the physical
processes which need to be treated in greater detail in order to
reproduce the observed spectra, and then iterating.
This is not generally practical, however.  Running a
simulation of the wind for even a single set of assumptions and
approximations takes a large amount of computer time and performing a large
number of iterations of simulating and comparing to observation is
generally impossible.  Furthermore, if a simulation compares
unfavorably to observational data, it may be unclear exactly how to
change the simulation in order to reproduce the observations.  Our
method of using simple velocity and density distributions, therefore,
serves as a complement to more computationally intensive calculations,
testing approximate wind densities and geometries to inform simulations.

Two-dimensional simulations of gravitational accretion from a smooth
flow onto a compact object using an adiabatic equation of state have shown that this
process is time variable and results in shocks (Fryxell \& Taam 1988;
Taam \& Fryxell 1989).  A series of simulations has considered the
effect of X-ray ionization on the wind from the OB companion.
Blondin et al.\ (1990) conducted two dimensional simulations of
accretion from a wind, including disruption of the radiative driving by
X-ray ionization of the wind, which showed the formation of accretion wakes
persisting over a significant fraction of the system size. They
suggest that the dense gas at the boundary of the accretion wake
could account for phase-dependent absorption features seen from HMXBs.
Blondin (1994) simulated HMXBs with large X-ray luminosities, including
the driving of a thermal wind from the X-ray irradiated face of the
companion star.  Only a few attempts have been made to
compare simulations with observational data quantitatively.  One
reason is that many simulations are done in two dimensions and the
quantitative extrapolation of the results to three dimensions is not
straightforward.

The wind in SMC~X-1 was simulated by Blondin \& Woo (1995) in three
dimensions, including UV radiation driving and its disruption by X-ray
ionization, and also the driving of a thermal wind from the X-ray
heated face of the companion.  This simulation included the gravity of
the supergiant companion and non-inertial forces but not the gravity
of the compact object.  The electron scattering and
fluorescence spectra from this simulation were calculated, and compared to {\it
Ginga} spectra by Woo et al.\ (1995).  The simulation was able to
reproduce the observed scattered continuum, as well as the extended
eclipse transition.  Wojdowski et al.\ (2000) observed
SMC~X-1 with {\it ASCA} over an eclipse, and computed the scattering,
fluorescence, and recombination from the simulation.  The observed
spectrum had no obvious emission features.  This excluded the
possibility that the DEM distribution of the wind contained a large
component with $1<\log\xi<3$, as strong recombination emission features
(predominantly O~VII and O~VIII) are emitted below 2 keV at those
ionization parameters.  In the simulation, the wind was denser in the
X-ray shadowed region,  but dense finger-like structures from the
shadowed side of the companion extended into the X-ray illuminated
side.  These structures were characterized by
ionization parameters in the range $1<\log\xi<3$, and their integrated emission
measure was greater than that allowed by the observed spectrum
by an order of magnitude.  It was therefore concluded that the
finger-like structures could not exist as they appeared in the
simulation. Again, it is not clear
how the physical assumptions of this simulation should be changed in
order to make the resultant wind conform to observations.

Despite the growing sophistication of these simulations, there are
still potentially important physical effects that have not been accounted for.
For example, none of the simulations mentioned above properly account for the
transfer of X rays through the
wind. To take an example, the inclusion of detailed
X-ray transfer may result in the formation of Str\"omgren
ionization boundaries around the X-ray source.  Outside of the 
Str\"omgren surface,
the wind is not highly ionized, thus UV radiation driving is not
strongly affected by the X-ray field of the compact object (Masai 1984).

\subsection{High-Resolution Spectroscopy and X-ray Line Profiles}

The high spectral resolving power of {\it Chandra\/} and {\it
XMM\/} bring the possibility of resolving line profiles.
Line widths of order 1000 km s$^{-1}$ have already been
observed with {\it Chandra} in Cir X-1 (Brandt \& Schulz 2000) and
Cyg X-3 (Paerels et al.\ 2000).  The canonical HMXB systems Cen X-3 and
Vela X-1, however, show emission lines with widths consistent with zero,
i.e., no broader than a few 100 km s$^{-1}$ (Wojdowski et al.\ 2001;
N.\ Schulz, priv.\ comm.).  For comparison, it is interesting to note
that  X-ray line widths
in isolated O and B type stars are as large as $\sim1000$ km s$^{-1}$
(Kahn et al.\ 2001; Schulz et al.\ 2000; Waldron \& Casinelli 2000).
This suggests one or both of the following possibilities:
(1) X-ray ionization slows the companion star
wind in HMXBs; (2) X-ray line emission from HMXBs traces a
different, lower velocity part of the wind than in isolated stars.

To illustrate the second possibility above, Fig.\ (3) shows the emissivity
contours of two lines, as calculated using the wind model of
Sako et al.\ (1999). The emissivities, which include the
$n_{\rm e}^2$ weighting, are by far the largest nearest the companion
star, in what for an isolated star would be the acceleration zone,
where the velocity is relatively low. This primarily shows that
outflowing material on the irradiated side of the companion is ionized to
high levels by the X-ray field long
before the wind attains a high velocity. Owing to this effect, X-ray 
emission line
profiles may often be fairly insensitive to wind velocity profiles.
It remains a puzzle as to why the  X-ray emitting plasma in the
Wolf-Rayet XRB system Cyg X-3 shows such large velocity widths.

\begin{figure}
\plotfiddle{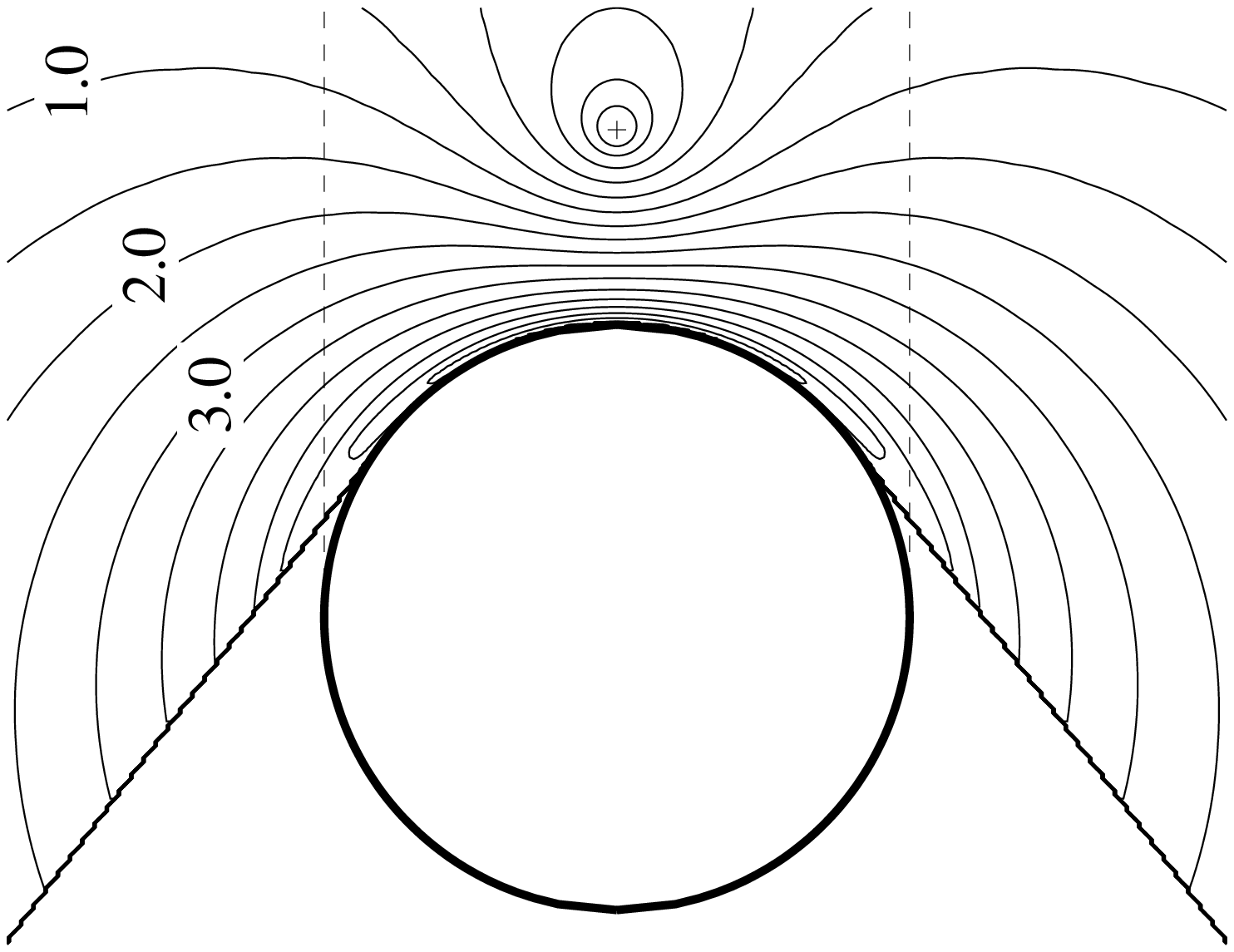}{2.9in}{-90.}{45.}{45.}{-310}{250}
\plotfiddle{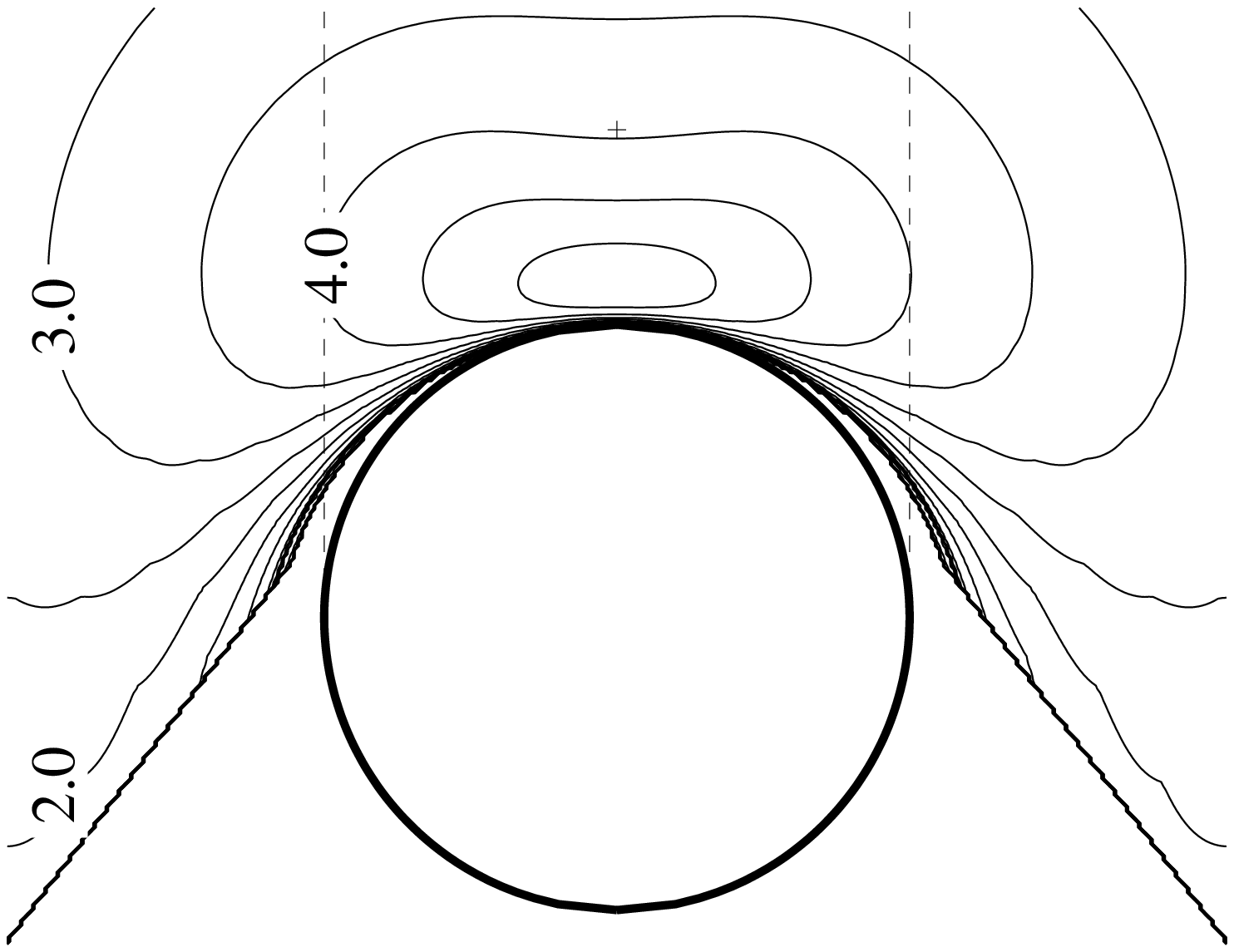}{2.9in}{-90.}{45.}{45.}{-110}{470}
\vskip -3.2in
\caption{Contour plots of the common logarithm of the line
emissivities (in ph\,cm$^{-3}$\,s$^{-1}$) of Si\,XIII ({\it left})
and Fe\,XXVI ({\it right}) assuming the Vela X-1 wind model used by
Sako et al.\ (1999). Position of the X-ray source is indicated
by a + sign.
Dashed lines delineate the part of the stellar wind occulted by
the companion star at orbital phase 0.0.
Diagonal lines tangent to the companion delineate the shadow cone.
Contours show the weighting of the line luminosities toward 
low-velocity regions
near the companion star.}
\end{figure}

When X-ray line optical depths are small, emission line widths and
shifts may be calculated by the methods described
in the previous section.  However, to simulate
situations where absorption is important (for example, systems in
which P Cygni lines have been observed) modifications must be made.
The emission spectrum for a given spatial cell of gas can still be treated as
though it were function of $\xi$ alone, but, in order to compute the 
total spectrum,
absorption along the line of sight to the observer must be applied to
each cell (see Wojdowski et al.\ 2000).

The {\it Chandra} X-ray spectrum of Circinus X-1 shows H-like and He-like lines
from Ne, Mg, Si, S, and Fe,
with line widths of $\sim 2000$ km s$^{-1}$ and P Cygni profiles,
the first observed in the X-ray band, indicative
of an accelerating outflow with a comparable velocity (Brandt \& Schulz 2000).
Interestingly, optical observations suggest that Cir X-1 is a low-mass system
(Glass 1994), which is also implied by its X-ray spectrum and
variability (Shirey, Bradt, \& Levine 1999), as well as its Type I X-ray bursts
(Tennant, Fabian, \& Shafer 1986).
Thus the X-ray continuum is thought to be viewed through an accretion 
disk wind,
which in some ways poses an even greater challenge for modelers
than stellar winds; for example, conditions at the base of the wind vary
with disk radius.
In order to obtain the values of $\xi$ corresponding to the
observed ionic species,
a density $n > 10^{14}$ cm$^{-3}$ is required.
This is consistent with the model
values for the atmospheric density in a hydrostatic disk (see \S5).
In addition, a broad ($\sim 2000$ km s$^{-1}$),  blueshifted
H$\alpha$ line may indicate an outflow of an optically thick gas,
indicative of even larger densities, perhaps at the base of the
wind (Johnston, Fender, \& Wu 1999).
A model spectrum of this source is not yet
available, but dynamic radiation-driven accretion disk wind
models have been made for AGN parameters (Proga, Stone, \& Kallman 2000).

\vskip 10pt
\noindent
Here we leave the topic of winds behind. In the following
sections, we outline some of the basic
issues relating to (``windless'') accretion disk structure, but note 
here that in
some situations, such as that exemplified by Cir X-1, the physics of
irradiated winds and the physics of irradiated accretion disks
cannot be so neatly separated.

\section{X-Ray Spectroscopy of Accretion Disks in LMXBs}

The intense emission from LMXBs is attributed to
the efficient conversion of gravitational potential
energy into radiation, as mass is transferred from an
otherwise normal star to a compact object (Shklovsky 1967).
The mass donor (companion) overflows its Roche lobe,
thereby spilling matter through the inner Lagrangian point (L1).
Born with large angular momentum in the co-rotating frame,
the accreting matter follows
trajectories that cannot directly intersect the compact object.
Instead, a disk forms, and a tangential viscosity forces the disk material
to spiral in (Prendergast \& Burbidge 1968).

Shakura \& Sunyaev (1973; SS73, hereafter) derived
an analytic model of radiatively efficient matter infall in a disk
geometry.  The viscously dissipated energy is assumed to be
locally radiated as blackbody emission. The UV continuum
of cataclysmic variables is in rough agreement with the above assumption
(Kiplinger 1979; Pringle, Verbunt, \& Wade 1986; Wade 1984).
The SS73 parameterization of the viscosity $\eta$ in terms
of the total pressure $P$, such that $\eta = \alpha P$, where $\alpha$
is a constant, provides a zeroth order picture of the disk, but it
is likely incorrect in detail. Numerical magnetohydrodynamic (MHD) 
models (see J.\
Stone article in this issue) show that $\alpha$ varies by a
few orders of magnitude throughout the disk, but that on average its
value ranges from $\sim 10^{-3}$--$10^{-1}$. Nevertheless, the $\alpha$-disk
model is one of the simplest and most useful of its kind, especially
since many results do not sensitively depend on the value of $\alpha$.
Comparisons of the
$\alpha$-disk model with observations of XRBs and AGN
are presented in Frank, King, \& Raine (1992) and Malkan (1983),
respectively.  Modern accretion flow models, of which the
$\alpha$-disk is a subset, are described by Chen et al.\ (1995).

The viscous mechanism in accretion disks allows accretion to occur by
transporting angular momentum outward, matter inwards, and by
dissipating gravitational energy into heat inside the disk. The viscosity
mechanism has been identified in theory as a magneto-rotational
instability due to the entanglement of magnetic fields from the
differential rotation of gas in Keplerian orbits, akin to a dynamo
effect (Balbus \& Hawley 1998). In MHD models, this mechanism 
provides the energy
dissipation and angular momentum transfer needed to naturally
produce mass accretion with self-sustained magnetic fields $B$ that
are much smaller than the equipartition level ($B^2/8\pi \ll \rho v^2$),
where $\rho$ is the gas density and $v$ is the thermal velocity.

The optically thick $\alpha$-disk model has been accepted with only a few
modifications since its inception
(Abramowicz et al.\ 1988; Chen \& Taam 1993; Luo \& Liang 1998),
and a significant fraction of
the X-ray continuum in LMXB is interpreted as thermal emission from
the disk.  Yet, the presence of non-thermal continuum radiation in all
of these sources and line emission from highly-ionized atomic species
in many of them is not accommodated by the standard $\alpha$-disk model.
Moreover, neutron star or coronal thermal
emission has made the discernment of disk continuum emission
ambiguous. As such, X-ray line emission is a potentially powerful tool for
elucidating the properties of accretion disks.

\subsection{Observations}

With the exception of Fe K emission in the 6.4--7.0 keV range,
discerning X-ray line emission in LMXBs has been
challenging, owing to limitations in sensitivity and spectral
resolving power, as well as the difficulties associated with
attempts to extract line emission from data dominated by
intense continuum emission.
Still, measurements obtained with the \it Einstein \rm Objective
Grating Spectrometer (Vrtilek et al.\ 1991),
the \it ROSAT \rm Position Sensitive Proportional
Counter (Schulz 1999), and the \it ASCA \rm CCD
imaging detectors (Asai et al.\ 2000) have shown
that the spectra of a large fraction of bright LMXBs
exhibit line emission.

The X-ray line emission arises presumably as
the result of irradiation of the disk by the X-ray continuum,
producing an extended source of reprocessed emission.
Evidence of X-ray emission from extended regions in LMXBs comes from 
the spectral
variations during ingress and egress phases of eclipses,
and during rapid intensity fluctuations known as dips.
Dips are thought to result from variable obscuration and attenuation
of the primary continuum by material near the outer disk
edge, which has been thickened due to impact of the
accretion stream with the disk (White \& Holt 1982;
cf., Frank, King, \& Lasota 1987). Hard X-ray emission, presumably
originating in an accretion disk corona
(ADC), and representing a few percent of the non-eclipse flux, remains
during mid-eclipse in several LMXB, implying that the ADC is larger
than the secondary star (White \& Holt 1982; McClintock et al.\ 1982). LMXB
spectra during eclipses or dips may harden or soften. Most sources
harden during dips (Parmar et al.\ 1986), consistent with
photoelectric absorption, but there are exceptions like the softening
of X1624-490 (Church \& Balucinska-Church 1995), and an unchanging X1755-338
(White et al.\ 1984; Church \& Baluncinska-Church 1993).
Sources such as X0748-676, X1916-053,
and X1254-690 show evidence for an unabsorbed spectral component
(Parmar et al.\ 1986; Church et al.\ 1997) during dips, revealing an
extended source of X rays which is larger than the ADC. These soft
X rays are likely radiation reprocessed in the accretion disk. Dip
ingress/egress times indicate ADC sizes in the $10^9$--$5 \times
10^{10}$ cm range, a factor of a few smaller than the accretion disk
sizes calculated from typical orbital parameters (Church 2000
and references therein).

The increasing availability of high-resolution X-ray spectra
from XRBs, thanks to {\it Chandra} and {\it XMM-Newton},
makes possible the application of detailed line diagnostics,
such as density-sensitive line ratios and line profiles.
Currently, however, only a few LMXB spectra from these observatories
are available. Of these, a 100 ks observation
of X0748-67 using the Reflection Grating Spectrometer (RGS)
shows the richest line spectrum (Cottam et al. 2000; Fig.\ 5)
and, therefore, provides a useful testbed for accretion disk atmosphere models.
System parameters of X0748-67 other than the orbital period
(3.82 hr) and inclination (75-82$\deg$) are unknown (Parmar et al.\ 1986).
The RGS lightcurve shows factor-of-10
fluctuations in flux, plus several Type I X-ray bursts with the
exponentially decaying flux expected from a nuclear detonation on the
surface of a neutron star. No correlation of the flux variability with
orbital phase was observed.
Excluding the burst phases,
spectra at low and high states were added for comparison.
The high state shows O VIII Ly$\alpha$ ($1s-2p_{1/2,3/2}$),
O VII He$\alpha$ ($n=2 \rightarrow n=1$; see \S2.1),
and Ne X Ly$\alpha$ line emission, with
O VII and O VIII absorption edges. The low state spectrum
shows even more lines but no edges, with the continuum emission
absent.  Emission from O VII and O VIII RRC can then be
detected, as well as Ne IX He$\alpha$ and N VII Ly$\alpha$.
The strongest line, O VIII Ly$\alpha$, is broader than the expected
line spread function, with a width of $\sigma =1390 \pm 80$
km/s. Upper limits for the O VII and Ne IX $R$ ratios
indicate a density $n_{\rm e} > 7 \times 10^{12}$
cm$^{-3}$. The line ratio $G > 4$, consistent with a purely photoionized plasma.
Unlike the hard X-ray flux, the RGS count rate does not
change during eclipses, which is indicative of an extended source.

\subsection{Radiatively Heated Accretion Disks}

In LMXB roughly half
of the gravitational potential energy is released in the vicinity of
the compact object (i.e., in accretion shocks near the neutron star
surface). The flared disk is exposed to this radiation, and will
be heated by it. In fact, radiative heating can exceed
internal viscous heating in the outer region of the disk.
The temperature structure of the disk can thus be controlled by the 
X-ray field,
photoionizing the gas, suppressing convection,  and increasing the scale
height of the disk. Photoionization in the disk is balanced by
recombination, which produces line emission.

Assuming that all the viscous heating and radiative heating from
illumination by the central source is
radiated locally as a blackbody (as in the SS73 model),
Vrtilek et al.\ 1990 find,
for a geometrically thin disk, and for $r \gg R_1$,
where $R_1$ is the radius of the compact X-ray source,
\begin{equation}
\sigma T_{\rm phot}^{4} \simeq \frac{3 G M_1 \dot{M} }{8 \pi r^{3} }
+ \frac{(1 - \eta) L_{\rm x} \sin \theta(r)}{4 \pi r^{2}},
\end{equation}
where $T_{\rm phot}$ is the photospheric temperature, $M_1$ is the mass
of the compact X-ray source, $\theta$ is the
grazing angle of the incident X-ray flux with respect to the disk surface,
and $\eta $ is the X-ray ``albedo,'' such that $1-\eta$ is the fraction of
X rays absorbed at the photosphere.
The second term on the right-hand side of Eq.\ (16) will be the
dominant one when
\begin{equation}
r>2.3 \times 10^8 ~\biggl(\frac{M_1}{M_{\sun}}\biggr) ~
\biggl(\frac{1 - \eta}{0.1}\biggr)^{-1}~
\biggl( \frac{\sin \theta}{0.1}\biggr)^{-1}~
\biggl(\frac{\epsilon_{\rm x}}{0.1}\biggr)^{-1} ~\rm cm,
\end{equation}
where we have written the X-ray luminosity in terms of an
``X-ray accretion efficiency'' $\epsilon_{\rm x}$, according to
$L_{\rm x}= \epsilon_{\rm x} \dot{M} c^2$. For example, accretion onto
a neutron star results in roughly 1/2 of the gravitational potential energy
being converted into X rays, or $\epsilon_{\rm x}
=GM_1/2c^2 R_1$.
The disk, therefore,
is radially divided into an inner region dominated by
internal dissipation, and an outer region dominated by external illumination.
External radiation will dominate the disk atmosphere energetics
for the outer two or three decades in radii, and the
local dissipation and magnetic flare heating, if any,
can be ignored there.

\subsection{Vertical Structure}

To obtain a high-resolution spectrum of an accretion disk, and in
particular one for which the outer layers are X-ray photoionized,
several authors have calculated the vertical structure by
solving the radiation transfer equations, assuming hydrostatic
equilibrium. Models have been applied to AGN and LMXB in the
high-$L_{\rm x}$ state, since in the low-state other accretion modes ensue. The
radiative transfer is typically simplified by using an
on-the spot approximation and the escape probability formalism.
Due to photoelectric absorption and Compton scattering, the
ionization structure of the disk turns out to be stratified,
approximated by a set of zones, each with a single ionization parameter.
The ionization structure of the disk can be solved by using
photoionization codes, such as CLOUDY and XSTAR,
to calculate the ionization and thermal equilibrium state
of the gas at each zone.

Ko \& Kallman (1991; 1994) calculated the vertical structure
of an illuminated accretion disk and obtained the
recombination X-ray spectrum for individual rings on the disk.
Raymond (1993) utilized the calculated temperatures in
Vrtilek et al.\ (1990) to calculate a vertical
structure and the UV spectrum from the global structure of the disk.
Both assumed parameters for LMXBs, and gas pressure-dominated disks.
Later models of photoionized accretion disks focused primarily on
calculating the Fe K$\alpha$ fluorescence emission from AGN disks.

Rozanska \& Czerny (1996) modeled semi-analytically the stratified, 
photoionized
transition region between the corona and the disk. They found that
their approximations, which included on-the-spot absorption, matched
more accurate radiation transfer codes, including optically thick
scattering, for optical depths $\la 10$. They also discussed the
existence of a two-phase medium, stopping short, however, of calculating
an X-ray line spectrum.
Nayakshin, Kazanas, \& Kallman (2000) modeled a radiation-pressure dominated
disk and showed that the vertical structure of the disk implied
significant differences in the Fe K fluorescence line spectrum compared to
that predicted by constant-density disk models
(e.g., Ross \& Fabian 1993; Matt, Fabian, \& Ross 1993; Zycki et al.\ 1994).
In addition, Nayakshin, Kazanas, \& Kallman (2000) found that the gas was
thermally unstable at certain ionization parameters, which created an
ambiguity in choosing solutions, and a sharp transition in temperature
in the disk.

Using the Raymond (1993) photoionization code,
we have calculated the disk atmosphere structure of each
annulus by integrating the hydrostatic balance and 1-D radiation
transfer equations for a slab geometry:
\begin{equation}
\label{eq:hydro}
\frac{\partial P}{\partial z} = - \frac{G M_{1} \rho z }{r^{3}}
\end{equation}
\begin{equation}
\label{eq:rad1}
\frac{\partial F_{\nu}}{\partial z} =
	- \frac{\kappa_{\nu} F_{\nu} }{\sin \theta}
\end{equation}
\begin{equation}
\label{eq:rad2}
\frac{\partial F_{\nu}^{\rm d}}{\partial z} =
	- \kappa_{\nu} F_{\nu}^{\rm d}
\end{equation}
where $P$ is the pressure, $\rho$ is the density,
$F_{\nu}$ is the incident radiation field, $F_{\nu}^{d}$ is the
reprocessed radiation propagating down towards the disk midplane,
$z$ the vertical distance from the midplane,
$\theta$ the grazing angle of the radiation on the disk,
$\kappa_{\nu}$ is the local absorption
coefficient. To these equations is added the further constraint
that local thermal equilibrium is satisfied:
\begin{equation}
\label{eq:thermal}
\Lambda (P, \rho,  F_{\nu}) = 0
\end{equation}
The difference between heating and cooling
$\Lambda$ includes Compton scattering, bremsstrahlung cooling,
photoionization heating, collisional line cooling, and recombination
cooling from H, He, C, N, O, Ne, Mg, Si, S, Ar, Ca, and Fe ions.
Equations (18)--(20) are integrated simultaneously using a
Runge-Kutta method with an adaptive stepsize control routine with error
estimation, and Eq.\ (21) is solved by a globally convergent
Newton's method (Press 1994). At the height $z_{\rm cor}$
(``cor'' for ``corona''), the equilibrium
$T$ is close to the Compton temperature,
from which we begin to integrate downward
until $T < T_{\rm phot}(r)$. The optically thick part of the disk, with
temperature $T_{\rm phot}$, is assumed to be vertically isothermal
(Vrtilek et al.\ 1990). We get $T_{\rm phot}$ from Eq.\ (16).

The boundary conditions at $z_{\rm cor}$ are set to
\begin{equation}
P(z_{\rm cor})= \rho_{\rm cor} k T_{\rm compton}/ \mu m_{\rm p}
\end{equation}
\begin{equation}
F_{\rm x}(z_{\rm cor})=L_{\rm x}/4 \pi r^2
\end{equation}
\begin{equation}
F_{\nu}^{\rm d}(z_{\rm cor})=0
\end{equation}
where $F_{\rm x} \equiv \int F_{v} ~d\nu$,
and $\mu$ is the average particle mass
in units of the proton mass $m_{\rm p}$. The boundary
conditions at $z_{\rm phot}$ for $F_{\nu}$ and $F_{\nu}^{\rm d}$ are set free,
and we use the shooting method (Press 1994)
with shooting parameter $\rho_{\rm cor}$, adjusted until
$P(z_{\rm phot})= \rho_{\rm phot} k T_{\rm phot} / \mu m_{\rm p}$
at the photosphere.
Note $\rho_{\rm phot}$ is the $\alpha$-dependent density
calculated for a standard SS73 disk.

An important feature of this model is that we allow the incident
radiation to modify the disk atmosphere geometry, such that
the heating and expansion of the atmosphere resulting from illumination
are used to calculate the height profile of the atmosphere
as a function of radius. The atmospheric height is
used to derive the input grazing angle of the radiation
for the next model iteration. This contrasts with calculating
the grazing angle using the pressure scale height
of the optically thick disk (Vrtilek et al.\ 1990), which is in general well
below the photoionized atmosphere, and which
underestimates the grazing angle and the line intensities
by an order of magnitude.
To get $T_{\rm phot}$ self-consistently from Eq.\ (16), we need
to find $\theta(r)$ using
\begin{equation}
\theta(r)= \arctan (dz_{\rm atm}/dr) - \arctan (z_{\rm atm}/r)
\end{equation}
where $z_{\rm atm}(r)$ is
defined as the height where the frequency-integrated grazing flux is 
attenuated by $e^{-1}$.
We stop after $\theta(r)$ and $T_{\rm phot}$ converge to $\la 10$\%.

\subsection{The Choice of Assumptions}

For modeling X-ray line emission from the disk atmosphere, the 
commonly used assumptions
of LTE and the diffusion approximation
will not hold. In addition,
assuming a constant density in the vertical direction will be inadequate,
since the hydrostatic equilibration time is small or comparable to
other relevant timescales, and the line emission is highly
sensitive to the vertical ionization structure. This is especially
true for recombination emission, in which each ion has a characteristic
set of line energies, and it contrasts with fluorescent line energies,
which vary little with ionization state until the atom is almost fully
ionized.
Yet, even fluorescence emission can be affected by a Compton-thick,
fully ionized gas above it (Nayakshin, Kazanas, \& Kallman 2000).

Hydrostatic equilibrium, thermal equilibrium, and ionization equilibrium
should be good assumptions in a time averaged sense.
The radiation transfer is complex,
and the assumptions used to simplify
calculations could be problematic.
In particular, by dividing the disk atmosphere into annular
zones with a given vertical gas column,
our 1-D radiation transfer calculation
assumes that 1) the primary continuum is not absorbed
before reaching the top of the column, 2) the
radiation in the column propagates from top to bottom at a given grazing
angle, and 3) there is no significant radiative coupling from one disk
annulus to another, which is used to justify the slab approximation.
The above assumptions are inadequate if the column height is comparable to
the disk radius, or if the photon mean free path in the gas column
is many times the local radius. Thus, future 2-D calculations
will result in better bookkeeping of photons, a more accurate structure,
and a more reliable X-ray line spectrum.

The correct calculation of line
transfer in the gas is also a concern, since the disk atmosphere is
optically thick in the lines. Line transfer is complicated by the
Keplerian velocity shear, which has to be taken into
account for a given viewing angle (Murray \& Chiang 1997).
The escape-probability approximation used to calculate line transfer
in the disk may also be inadequate because of the large optical depths.

Our calculations show that the proper treatment of a thermal
instability (Field 1965; Krolik, McKee, \& Tarter 1981) and the 
effect of conduction to
resolve it affect the spectrum significantly (Zeldovich \& Pikelner 1969).  A
two-phase gas could form, with clouds of an unknown size distribution
and with undetermined dynamics of evaporation and
condensation (Begelman \& McKee 1990), with each phase having distinct
ionization parameter and opacity. The instability is sensitive to 1) the metal
abundances, 2) the continuum shape (Hess, Kahn, \& Paerels 1997), and 3) the
atomic kinetics (Savin et al.\ 1999). Finally, the inclusion of 
radiation pressure
in high luminosity systems or for small radii is advisable
(Proga, Stone, \& Kallman 2000).

The local viscous energy dissipation rate per unit volume in the disk
atmosphere can be included in Eq.\ (21) with the form
(SS73; Czerny \& King 1989)
\begin{equation}
Q_{visc} = \frac{3}{2} \Omega \alpha P
\end{equation}
where $\Omega$ is the Keplerian angular velocity, $\alpha$ is the
viscosity parameter, and $P$ is the local total gas pressure.
Equation (26) is an extension of the $\alpha$-disk model,
where the viscous dissipation is vertically averaged,
assumes the local validity of the $\alpha$
prescription, and is untested. Fortunately, our numerical modeling
indicates that the viscosity term is negligible in most regions of the
disk atmosphere except for the inner disk if $\alpha \sim 1$ (in particular,
near the Compton-temperature corona). Its effect is to enhance
a thermal instability between $10^6$ and $10^7$ K.  Vertically
stratified MHD models (Miller \& Stone 2000), although inconclusive, owing to
the uncertain effect of boundary conditions, show that
the viscous dissipation drops rapidly $\ga 2$ pressure scale
heights away from the disk midplane, providing evidence against
Eq.\ (26). The disk atmosphere is always a few scale heights
above the midplane. Therefore, we choose not to include
this term in our models.  Equation (26) has been applied in
the optically thick regions of the disk (Dubus et al.\ 1999) and in the disk
atmosphere (Rozanska \& Czerny 1996). Other forms for the local
dissipation that reduce to the $\alpha$-disk have been used
(Meyer \& Meyer-Hofmeister 1982).

\subsection{Thermal Instability}

Irradiated gas is subject to a
thermal instability for temperatures in the $10^5$--$10^6$ K range
(Buff \& McCray 1974; Field, Goldsmith, \& Habing 1969),
suppressing X-ray line emission in that regime.
The Field (1965) stability criterion indicates that
a photoionized gas may become unstable when recombination
cooling of H-like and He-like ions is important.
The temperature and ionization parameter ranges where the instability
occurs depend on the metal abundances and the local radiation spectrum
(Hess, Kahn, \& Paerels 1997).
Within a range of {\it pressure ionization parameters}
($\Xi \equiv P_{\rm rad}/P_{\rm gas}$, where $P_{\rm rad}$ is
the radiation pressure, and $P_{\rm gas}$ the gas pressure; Krolik, 
McKee, \& Tarter 1981),
thermal equilibrium is achieved by three distinct temperatures, only
two of which are stable to perturbations in $T$ (see Fig.\ 4).
The instability implies a large temperature gradient
as the gas is forced to move between stable branches, requiring
the formation of a transition region whose size could be determined by
electron heat conduction.

\begin{figure}
\plotone{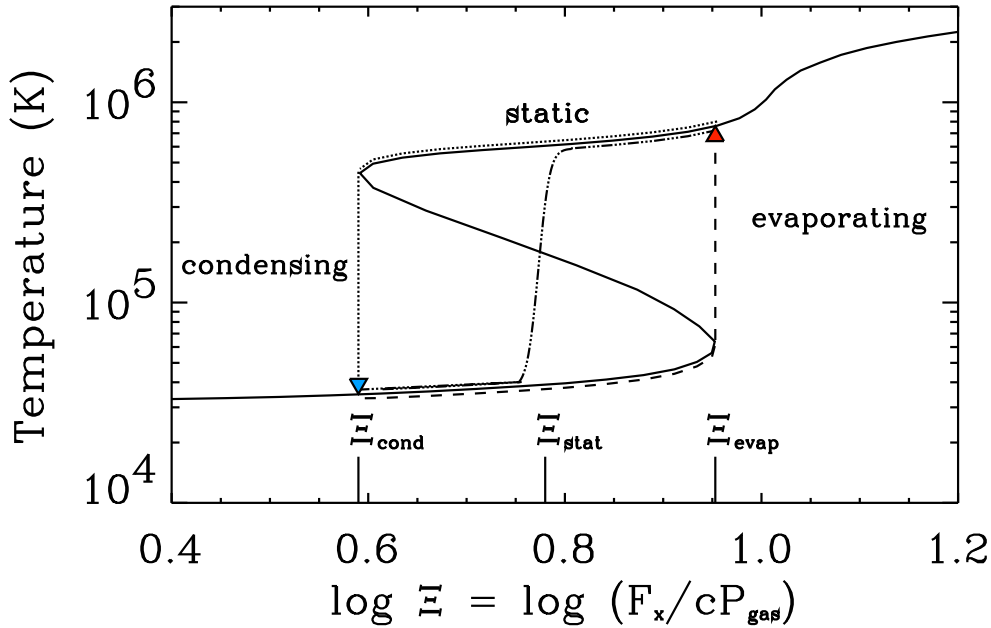}
\caption{Electron temperature vs.\ pressure ionization parameter,
assuming an 8 keV bremsstrahlung for the spectral shape of the
irradiating continuum. The solid $S$-curve
corresponds to the locus of solutions of the equations of thermal balance
and ionization equilibrium. Stable solutions have positive slope.
We model the extreme cases, the evaporating and
condensing disks, which correspond to the lower and upper stable branches
of the curve, respectively. The inclusion of electron conduction modifies
the $S$-curve to the dotted, dashed, or dash-dotted curves.
The dash-dotted curve is a schematic of the static solution,
the dotted curve corresponds to the extreme condensing solution,
and the dashed curve to the extreme evaporating solution.
Above $\Xi=1.2$, the equilibrium temperature rises to the Compton
temperature $\sim 10 ^7$ K (not shown). }
\end{figure}

For simplicity, we neglect emission from the conduction transition region.
Upon calculation of the Field length,
the length scale below which conduction dominates
thermal equilibrium, we estimate that conduction
forms a transition layer $\sim 10^{-2}$ times the geometrical
thickness of the X-ray emitting zones. Nevertheless, X-ray
line emission from the neglected transition region could be important
(Li, Gu, \& Kahn 2001).

Consideration of conduction gives a physical interpretation to
choosing one of the two stable solutions for a given $\Xi$.
Roughly, a static conduction solution must split the instability
region in half, taking the low-$T$ stable branch
at $\Xi < \Xi_{\rm stat}$, and the high-$T$ stable branch at $\Xi >
\Xi_{\rm stat}$, separated by the transition layer at
$\Xi_{\rm stat}$ (Zeldovich \& Pikelner 1969).
A transition layer located away from $\Xi_{\rm stat}$ will dynamically approach
$\Xi_{\rm stat}$ by mass flow. Thus, a transition layer at
$\Xi_{\rm evap}>\Xi_{\rm stat}$ corresponds to evaporating gas, while a
transition at $\Xi_{\rm cond}<\Xi_{\rm stat}$ implies condensing gas
(Zeldovich \& Pikelner 1969).

We compute the disk structure for both condensing
and evaporating solutions (Jimenez-Garate et al.\ 2001).
The static conduction solution is an
intermediate case of the latter extreme cases.
We always take a single-valued $T(\Xi)$, since a two-phase solution may be
buoyantly unstable, making the denser, colder gas sink.
The evaporating disk corresponds to the low-$T$ branch of the 
instability, while
the condensing disk corresponds to the high-$T$ branch.

\subsection{Preliminary Comparisons with Observed Spectra}

Using a model for a disk illuminated by a neutron star, we
have produced an X-ray recombination line spectrum from a
self-consistent accretion disk atmosphere. The flaring of the atmosphere
is consistent with optical observations (de Jong, van Paradijs, \& 
Augusteijn 1996).
By introducing the feedback between illumination and disk atmosphere
geometry (see Eq.\ 25), we have obtained a disk with
a height to radius ($z_{\rm atm}/r$) ratio that is $\sim10$ times larger
than previously thought, with a corresponding increase in the
reprocessed line radiation intensity.

With the disk structure [$\rho(r,z)$, $T(r,z)$]
and charge state distributions $f_{Z,i}(r,z)$,
we compute local X-ray emissivities using detailed atomic models.
The atomic structure, radiative transition rates, and collisional 
rate coefficients are
calculated with HULLAC (Hebrew University/Lawrence
Livermore Atomic Code; Klapisch et al.\ 1977).
Radiative recombination rates, including the RRC, are found from
the Milne relation using the photoionization cross-sections provided by
Saloman, Hubble, \& Scofield (1988). The overall spectral model includes
the H-like and He-like ions of C, N,
O, Ne, Mg, Si, S, Ar, Ca, and Fe, as well as the Fe L-shell ions.
Further details are described in Sako et al.\ (1999).

Each annulus consists of a grid of zones in the vertical
direction, and $T$, $\rho$ and $f_{Z,i+1}$ for each zone
are used to calculate the line and RRC emissivities. The radiation
is propagated outwards at inclination angle $i$, including
the continuum opacity of all zones above.
The spectrum is Doppler broadened by the projected
local Keplerian velocity, assuming azimuthal symmetry.
The outgoing spectrum
for each annulus is added to a running total to obtain the total disk
spectrum.

Our models show that, for a low-inclination LMXB,
X-ray lines from the disk will be difficult to detect against
the bright continuum.
The \it Chandra \rm LETGS  spectrum of X0614+091 (Paerels et al.\ 2001)
confirms this. Although a number of interstellar
absorption features are present, there is no evidence for disk line emission.
The data can be reproduced, 
i.e., we can generate a model spectrum
for which the X-ray line emission lies below the limit
of detectability, if we constrain the
outer disk radius such that $r \la 10^{10}$ cm.
We propose that this source is at a low
inclination and has a small accretion disk, reducing substantially
the equivalent widths of disk emission lines. Unfortunately, the 
orbital parameters and
inclination of this source are unknown. Of course, this is only
a consistency check---not
the best way to corroborate details of disk models---but it is
still somewhat gratifying that there are plausible pockets of parameter space
for which the model predicts unobservably weak lines.

\begin{figure}
\plotone{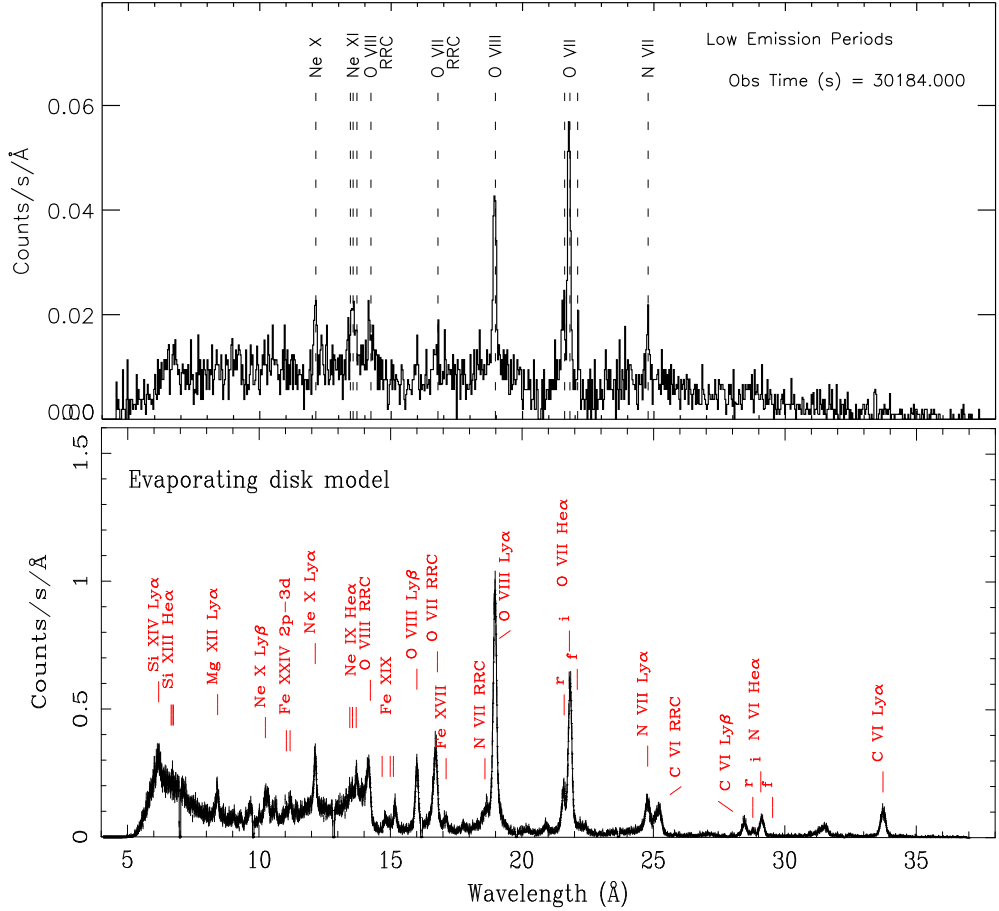}
\caption{\it Top: \rm Spectrum of X0748-67 observed with the \it
XMM-Newton \rm RGS during 30 ks of low (dipping) states (from Cottam
et al.\ 2000).
\it Bottom: \rm Simulated spectrum of a 50 ks observation for an
Eddington luminosity LMXB at 10 kpc, with $N_{H} = 5 \times 10^{22}$
cm$^{-2}$ for the central flux, and $N_{H} = 10^{21}$ cm$^{-2}$ for
the disk emission.}
\end{figure}

The \it XMM-Newton \rm RGS spectrum of X0748-67, which shows
a number of emission lines (\S5.1),
provides a more stringent test. We have not yet attempted
to ``fit'' the data; the particular result discussed here arises from 
a preliminary
investigation of the sensitivity of the X-ray line spectrum to
parameter variation and selecting
among different thermally stable solutions
(Jimenez-Garate et al., in preparation).
As can be seen from  (Fig.\ 5),
our disk atmosphere model spectrum compares favorably with the X0748-67 data.
The model disk inclination is set to $75\deg$.
For the line emission, we applied a neutral absorbing
column density $N_{H}= 10^{21}$ cm$^{-2}$, and
for the continuum emission $N_{H}= 5 \times 10^{22}$ cm$^{-2}$,
which represents a geometry for which a
compact continuum source is
obscured by material in the outer disk. This is
consistent with the significant variability in the continuum,
and the absence of variability in
the extended disk line emission.
Integration of theoretical He-like spectra over the disk
produce He-like $R$ ratios that are consistent with those observed.
For example, the typical model atmospheric density corresponding to
O VII emission is $n_{e} \sim
10^{14}$ cm$^{-3}$ or greater. Hydrogen-like O VIII RRC are very broad,
with a superposed narrow component in evaporating disks.
The O VII RRC and the $w$ line appear at the expected relative levels.
Among other preliminary results, we find that the
O VIII/O VII line ratios
are good tracers of evaporation from the disk (Jimenez-Garate et al.\ 2001),
so that the predicted O VIII Ly$\alpha$ to O VII He$\alpha$ ratio, for example,
is sensitive to both the flaring angle of
the disk and to the presence or absence of
evaporation. The robustness of the calculated
ratio against changes in the computational methods
used will have to be evaluated {\it vis-a-vis}
other models, e.g., those using 2-D radiation transfer.
The O VII He$\alpha$ line flux may be more model
dependent since it is produced in a smaller region,
where conduction is not negligible.

\begin{figure}
\plotone{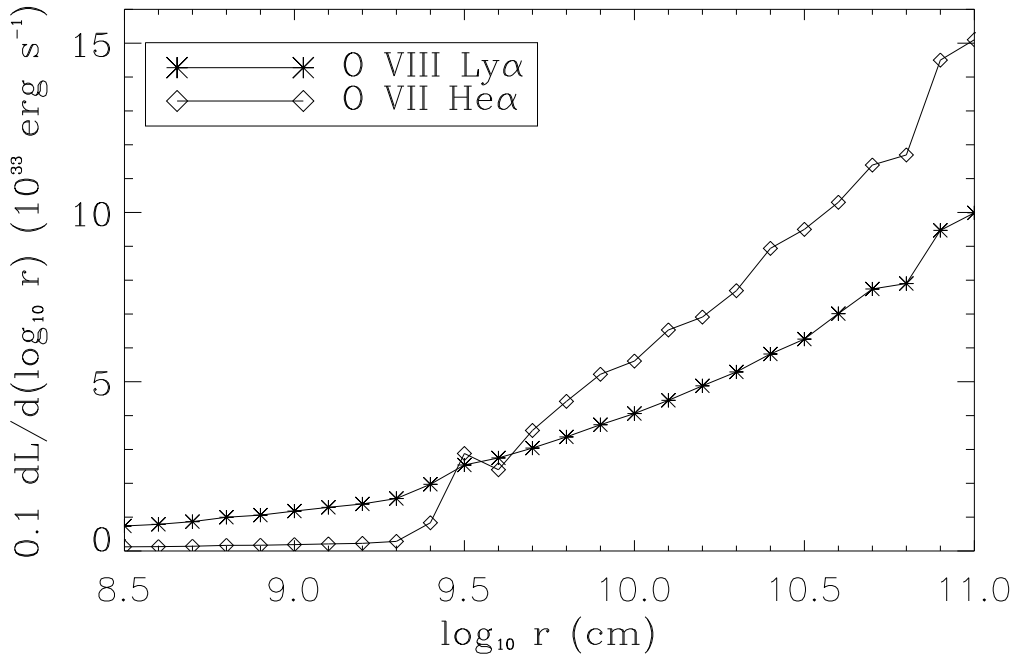}
\caption{Luminosity per bin as a function of radius of
the lines from H-like and He-like oxygen for the disk atmosphere
model described in \S5.3. 
Each line is emitted throughout a large region of the disk.}
\end{figure}

Line profiles for X0748-67are resolvable by the \it XMM \rm and
\it Chandra \rm spectrometers. We find that an essential factor
in properly calculating line profiles is that X-ray line formation,
for a given line,
is spread over most of the disk. For example, the $dL/dr$ profile of 
O VIII Ly$\alpha$
(Fig.\ 6) shows that radii spanning several orders of magnitude
($10^{8.5} \leq r \leq 10^{11}$ cm)
need to be accounted for (this applies also, of course, to a proper
calculation of the total line luminosity or line equivalent width).
The theoretical line profile for a single narrow annulus
is double peaked. However, integrated over the disk surface, the
line acquires broad, smooth wings and only a very narrow double-
peaked core from the outermost annulus. The spectrometer response
can easily smooth this feature into a quasi-Gaussian core profile.
Again we see that
access to a global, parameterized model of the source is imperative if
we are to provide a valid interpretation---not just a description---of
spectroscopic data.
The erstwhile goal of
inferring accretion disk structure by working backwards from a set
of plasma diagnostics appears to be an untenable proposition.

\section{Final Comments}

We have remarked that, thus far, X-ray emission line
spectra of XRBs consist almost exclusively of H-like
and He-like ion spectra, mixed with K$\alpha$ fluorescence lines. While the
atomic physics and population kinetics of the former
are fairly simple, the latter are, in principle, quite
complicated, but merely {\it treated} as though they were simple. To 
some extent,
this situation removes the atomic physics barriers to successful 
interpretations of X-ray spectra from XRBs.
With increasing detector areas, however, such as will
become available with {\it Constellation-X},
more advanced techniques will become more readily available (Fe 
L-shell spectroscopy,
ion fluorescence spectroscopy, etc.). At that time, atomic
physics issues and uncertainties will again be a primary concern.
For the time being,
it is our opinion that there are more pressing problems.

As has been discussed,
line formation in an XRB is inherently global, in the sense
that line luminosities typically involve integrations over
non-negligible fractions of the source volumes, over which the basic
physical parameters may assume a broad range of values.
Although it is useful to derive semi-quantitative
estimates of densities, temperatures, and velocities from
observed spectra, the notion that a particular value
of a particular diagnostic line ratio allows us to specify
one of these physical parameters fails to acknowledge
this global nature. What is needed, therefore, is development of and access
to global models of XRBs---models that strive for spectroscopic 
accuracy, so that
X-ray spectra now being obtained with {\it Chandra} and {\it 
XMM-Newton} can be used
as feedback. We have described a few of these models.
Those that have been successful at reproducing X-ray spectra are full
of simplifying assumptions, thereby lacking sufficient physical realism.
Unfortunately, adding complexity to models does not guarantee their success,
owing to the need to select among various subsets of physical assumptions.
In the long term, radiative-hydrodynamic calculations
will undoubtedly provide an increasingly realistic picture of accretion
flows in XRBs. In the near term, however, we are faced with a number 
of obstacles, both conceptual and practical, that remain to be surmounted.

\acknowledgments
We would like to acknowledge contributions to this paper
and the accompanying presentation by
Jean Cottam, Julia Lee, Christopher Mauche, Koji Mukai, John Raymond, 
and Norbert Schulz.
D.A.L.\ was supported in part by the NASA Long Term Space Astrophysics Program
grant S-92654-F. Work at LLNL was performed under the auspices of the U.S.\
Department of Energy by the University of California
Lawrence Livermore National Laboratory under contract No.\ W-7405-Eng-48.

\end{document}